# SNIFFLES – I. Intended emission, unwanted emission, and unintended radiation from low-Earth orbiting satellites impacting radio astronomy from 1 to 26 GHz

Balthasar Indermuehle ★ and Liroy Lourenço
*CSIRO Space & Astronomy, PO Box 76, Epping, NSW 1710, Australia*

## ABSTRACT

We present the first results of SNIFFLES, an ongoing observational programme to characterize intended emission, unwanted emission, and unintended radiation from non-geostationary satellite orbit (NGSO) systems across common radio astronomy receiver bands from 1 to 26 GHz (*L*, *S*, *C*, *X*, and *K* bands). Using the Australian 22 m Mopra radio telescope near Coonabarabran, New South Wales, with follow-up observations from a single dish of the Australia Telescope Compact Array (ATCA) interferometer and its new BIGCAT backend, we conducted 4629 tracked observations of satellites from four NGSO constellations (Starlink, OneWeb, Amazon Leo, and Guowang), amounting to 375.9 h of telescope time. Satellite identification was confirmed by correlating detected Doppler shifts with predicted ephemerides. We detected 2345 instances of intended emission, unwanted emission, and unintended radiation at 300+ unique frequencies from three of the four systems. The detections span all three interference classes affecting radio astronomy: (1) intended emission [including Direct-to-Cell (DTC)], (2) unwanted emission (out-of-band emission, including up to the fourth harmonic of the Starlink DTC signal at approximately 2.6 GHz, with the fourth harmonic detected near 10.5 GHz), and (3) unintended radiation from satellite platform electronics. Several detections fall within primary radio astronomy allocations, including 1613.19 MHz within the protected OH line band, and at 2690.76 and 2700 MHz. At 2700 MHz, unintended radiation was detected in 76.9 per cent of all observations of the relevant satellite version. ATCA follow-up measurements confirm flux densities up to 11 orders of magnitude brighter than typical astronomical sources, well in excess of levels that saturate radio astronomy receivers.

## 1 INTRODUCTION

Large satellite constellations are increasingly being deployed in low- and medium-Earth orbit (LEO/MEO), providing ubiquitous connectivity across the entire globe. The effects of their intended emissions, unwanted emissions, and unintended radiation from satellite platform electronics (e.g. solar power inverters, computers, and networking equipment) have been shown to significantly impact terrestrial radio astronomy. Previous studies by F. Di Vruno et al. (2023), D. Grigg et al. (2023), C. G. Bassa et al. (2024), D. Grigg et al. (2025), and X. Zhang et al. (2025) have all detected such unintended radiation with different radio telescope instruments in Europe and Australia. To describe this new kind of interference, a new term was coined in these early papers: unintended electromagnetic radiation (UEMR). The aforementioned studies found UEMR-class unintended radiation at low frequencies that are particularly troublesome for low-frequency beamforming arrays such as LOw-Frequency ARray (LOFAR; M. P. van Haarlem et al. 2013), the Long Wavelength Array (S. W. Ellingson et al. 2009), or the low-frequency component of the SKA[1] currently being built in Australia. These arrays typically operate in the frequency band from 50 to 350 MHz (R. Ekers 2012), a range that encompasses some of the highest priority science targets in modern astrophysics, including the redshifted 21 cm hydrogen signal from the Cosmic Dawn and the Epoch of Reionization (M. F. Morales & J. S. B. Wyithe 2010). Experiments such as EDGES (J. D. Bowman et al. 2018) are seeking to detect this faint cosmological signal, which requires an exceptionally clean radio-frequency environment. The significant power levels of the unintended radiation and unwanted emissions – often hundreds to thousands of Janskys ($10^{-24}$ to $10^{-23}$ W m$^{-2}$ Hz$^{-1}$) – prompted the authors of this study to look for unintended radiation and unwanted emissions at higher frequencies. To align our work better with existing terminology from the International Telecommunications Union–Radiocommunications Sector (ITU-R) Radio Regulations (ITU-R 2025), we will refer to UEMR as unintended radiation. We will also provide background on the unusual sounding but very narrowly defined terms we use in Section 2.3.

★ E-mail: balt.indermuehle@csiro.au

[1] https://www.skao.int/en/explore/telescopes/ska-low

The observations presented in this paper form the first results of SNIFFLES: Satellite measuremeNt of Intended emission, unwanted emission, and radio radiation to develop, Follow up, and veriFy mitigation measures, regulatory compliance, and Lawful usE of the radio Spectrum. SNIFFLES is an ongoing monitoring programme designed to systematically characterize the electromagnetic environment created by non-geostationary satellite orbit (NGSO) constellations across the suite of Australia Telescope National Facility (ATNF) receivers. Specifically, we extended the investigation of intended emission, unwanted emission, and unintended radiation into higher frequency bands in regular use by radio astronomy. The bands covered in our work span 1–26 GHz (*L*, *S*, *C*, *X*, and *K* bands). The bulk of observations were obtained using the Mopra telescope near Coonabarabran, New South Wales, with additional verification observations conducted using a single dish of the Australia Telescope Compact Array (ATCA) interferometer in Narrabri.

We find that as the orbital density of satellites has reached unprecedented levels, their intended emissions, unwanted emissions, and unintended radiation are impacting radio astronomy observations to a significant degree. Beam coupling between the satellite transmitter main beam and the radio astronomy antenna main lobe or side lobes is also becoming an increasingly common occurrence. Although radio astronomers try to actively avoid pointing the sensitive radio astronomy receivers at satellites, modelling shows that such avoidance will increasingly become operationally impossible. Based on satellite system filings at the ITU-R for the SpaceX Starlink Gen 2, OneWeb phase 2, and GuoWang constellations alone (noting that not all filings will necessarily result in launches), Indermuehle (2024) forecasts that beam coupling between a radio astronomy station and at least one of these satellites will occur every few minutes, assuming half-degree beam widths for the main beam of the radio astronomy telescope. In some regions of the sky the rate will be even higher, depending on the orbital geometries of the constellations. This results in inadvertent high-gain observations of the satellite platform electronics and their transmitters. Even without beam coupling, power received from the intended emissions through the telescope sidelobes is strong enough to remain clearly visible in the radio astronomy receiver bands across the radio horizon of the satellites. When beam coupling does occur, the powerful intended emissions can overpower the low-noise amplifiers in the radio astronomy receivers, resulting in a non-linear response. Samplers can also become overwhelmed with signal levels in the mega-Jansky range ($> 10^{-20}\,\mathrm{W\,m^{-2}\,Hz^{-1}}$), rendering the entire bandpass unusable, forcing the data to be discarded. This can lead to a projected data loss for radio astronomy of the order of 15 per cent in the affected bands (Indermuehle 2024).

For these reasons, we have elected to undertake observations of satellite platforms, not just in the bands protected for the radio astronomy service (RAS) in the Radio Regulations (RRs; ITU-R 2024), but throughout the bands where radio astronomy needs to be able to observe the universe to meet its science goals. We remain hopeful that studies like this will help in obtaining better protection for radio astronomy through regulatory and voluntary measures by the satellite operators.

The paper is structured as follows. Section 2 outlines the regulatory landscape that we consider relevant to unintended radiation from NGSO satellites. Section 3 describes the satellite systems examined in this study and the observing methods employed, including both Mopra's stepped-band survey and ATCA's high-resolution follow-up measurements. Section 4 presents the detection results across all bands and constellations. Finally, Section 5 concludes with the implications of these findings for radio astronomy and scopes future work.

## 2 REGULATORY LANDSCAPE

### 2.1 ITU-R

The radio spectrum is allocated to services by way of an international treaty: the International Telecommunications Union – Radiocommunications Sector (ITU-R) RRs (ITU-R 2024). The RRs are an international treaty framework negotiated approximately every 4 yr at the World Radio Conference (WRC) – the next one to be held in 2027. ITU-R is also responsible for satellite orbit filings, so any satellite operators wishing to launch a satellite constellation will be required to file this with ITU-R, and have a window of 7 yr from the date of filing until a first satellite must be in orbit. As of November 2025, informal communication from the ITU-R's satellite filing branch indicated that 1.8 million satellite filings had been received, notionally to be launched over the next decade. In addition to the ITU filings, and noting this is separate from ITU filings, SpaceX have filed a regulatory concept constellation of 1 million orbital data centre satellites with the United States Federal Communications Commission (FCC) at the end of January 2026 (FCC 2026).

As a specialized agency of the United Nations, ITU-R is a contribution-driven assembly of 194 member states (referred to as administrations), and close to 1000 sector members and academic institutions (ITU 2024). While sector members and academia do not have a vote at ITU-R, they can participate in the relevant Study Groups (SGs) and Working Parties and provide input documents and participate in their work. Examples of sector members include telecommunications providers, satellite companies, equipment manufacturers, and scientific organizations with a vested interest in the use of the radio spectrum.

### 2.2 Definitions of emission categories used in this work

Throughout this paper, we classify detections into three categories: intended emission, unwanted emission, and unintended radiation. Because this taxonomy underpins much of the analysis that follows, and because the third term has no formal standing in the RRs, we define our usage explicitly here before discussing the regulatory framework that applies to each category.

#### *2.2.1 ITU-R definitions*

The RRs define *emission* (RR 1.138) as 'radiation produced, or the production of radiation, by a radio transmitting station', where radiation (RR 1.137) is 'the outward flow of energy from any source in the form of radio waves'. Emission is therefore a strict subset of radiation: every emission is radiation, but not all radiation is emission. The RRs regulate emission directly, while radiation from sources other than a transmitting station is addressed only indirectly, principally through RR 15.12.

Emissions occurring outside the necessary bandwidth of an authorized transmission are termed *unwanted emissions* (RR 1.146) and comprise two subcategories:

(i) **Out-of-band emissions** [out-of-band emission (OOBE), RR 1.146A]: Emission on frequencies immediately outside the

necessary bandwidth that result from the modulation process, excluding spurious emissions.

(ii) **Spurious emissions** (RR 1.145): Emissions on frequencies outside the necessary bandwidth whose level can be reduced without affecting the transmission of information. This category explicitly includes harmonic emissions, parasitic emissions, intermodulation products, and frequency conversion products.

#### *2.2.2 Where unintended radiation sits relative to RR-defined emissions*

Neither subcategory of unwanted emission cleanly accommodates radiation produced by satellite platform subsystems that are not part of the radio transmitter chain, for example, switched-mode power converters in the spacecraft bus, on-board computing and networking equipment, reaction-wheel drive electronics, or solar-array power conditioning. Such radiation has no necessary bandwidth to be 'outside of', is not produced by a modulation process, and is not in general traceable to any specific intended carrier. It is, however, unambiguously radiation in the sense of RR 1.137, and as radiation from electrical apparatus aboard a station it falls within the scope of the obligations set out in RR 15.12.

To distinguish this third class from the RR-defined unwanted emissions, we use the term *unintended radiation*, which is co-extensive with the UEMR terminology adopted by F. Di Vruno et al. (2023) and subsequent authors. We prefer *unintended radiation* on the grounds of closer alignment with existing RR vocabulary; the two terms refer to the same physical phenomenon. The three categories used in this paper therefore map on to the following:

(i) **Intended emission**: Radiation within the authorized transmission of a licensed carrier (or one operating under RR 4.4), including the fundamental signal of the operator's intended service.

(ii) **Unwanted emission**: OOBE and spurious emissions as defined in RR 1.146, including harmonics, intermodulation products, and modulation sidebands of an intended carrier.

(iii) **Unintended radiation**: Radiation from satellite platform subsystems not associated with any authorized carrier, including but not limited to emissions from on-board electronics, power conditioning, and digital subsystems.

#### *2.2.3 Assignment criteria for detections*

Each detection in our data base was assigned to one of the three categories above using the following decision rule, applied in order:

(i) **Intended emission:** A detection is classified as an intended emission if its centre frequency falls within an allocation that the satellite operator is licensed, or otherwise permitted under RR 4.4, to use for that satellite generation. The Starlink Direct-to-Cell (DTC) downlink at 2620–2630 MHz (Fig. 4) is the canonical example.

(ii) **Unwanted emission:** A detection is classified as unwanted emission if its centre frequency lies at an integer harmonic, subharmonic, intermodulation product, or modulation sideband of a known intended carrier of the same operator, within a tolerance set by the carrier bandwidth or spectral features, and the expected Doppler shift. The detection at 5240–5260 MHz (Fig. 7a and Table 5) is classified as the second harmonic of the 2620–2630 MHz DTC downlink because: (i) it is centred at 2× the DTC carrier frequency; (ii) its bandwidth and spectral features match those of the fundamental emission extrapolated to the second-order harmonic, (iii) its observed Doppler rate is 2× that of the DTC fundamental, consistent with a harmonic of the same emitter; and (iv) it is observed only on DTC-enabled satellites, and only on passes during which those satellites are actively transmitting in the DTC band. The third and fourth harmonics reported in Table 5 satisfy the same criteria.

(iii) **Unintended radiation:** A detection is classified as unintended radiation if it cannot be associated with any known intended carrier of the operator under criteria (1) or (2). Detections that are persistent across a satellite version but bear no integer-ratio or sideband relationship to any operator-authorized frequency, such as the 2700 MHz signal observed on 76.9 per cent of Starlink V2-Mini Ku-optimized passes (Table A3), or the cluster of narrow lines near 2660 MHz, are placed in this category. In some instances, we also have confirmation from the satellite operator that they have measured radiation being produced at that frequency in anechoic chamber measurements in their efforts to understand and reduce unintended radiation.

The Doppler cross-check in criterion (2) is particularly useful for resolving ambiguous cases. A true $n$th harmonic of a transmitted carrier exhibits a Doppler shift that is exactly $n$ times that of the fundamental, whereas radiation from an independent on-board oscillator that happens to lie near $n\times$ a transmitted frequency will exhibit a Doppler shift consistent with 1× the apparent radial velocity of the satellite. A small number of detections in our data base remain ambiguous under this rule, primarily where a candidate harmonic frequency coincides with a frequency at which platform radiation has been independently detected on non-transmitting satellites of the same generation; these cases are flagged as such in the relevant tables and counted conservatively as unintended radiation pending further follow-up with ATCA/BIGCAT.

### 2.3 Relevant principles in the radio regulations

In this section, we present a cohesive argument as to why unwanted emissions and unintended radiation falling within RAS allocated bands should be regulated through the existing RR framework. We note that minimizing unintended radiation from satellite platform electronics is also a worthwhile goal beyond the RAS bands, but acknowledge that the regulatory basis for addressing radiation outside allocated bands is less well established.

In addition to the terminology defined in Section 2.2.1, the RRs have clear definitions for the terms 'station' and 'space station', defined in RRs 1.61 and 1.64, respectively. We propose that these definitions encompass all electromagnetic radiation from the space stations: The definition of 'station' in RR 1.61 states '[...] including the accessory equipment, necessary at one location for carrying on a radiocommunication service [...]'. These definition inclusions would make little sense if there was no intent to regulate radiation as defined in RR 1.137 from non-transmitting equipment.

RR 15.12 clearly establishes the regulatory mandate of ITU-R beyond radio transmitter equipment: 'Administrations shall take all practicable and necessary steps to ensure that the operation of electrical apparatus or installations of any kind, including power and telecommunication distribution networks [...] does not cause harmful interference to a radiocommunication service[...]'. Satellite systems providing global communication coverage are a telecommunication distribution network.

This furthermore has implications for the obligation to avoid harmful interference as laid out in Preamble 0.4 of the RRs: 'All stations, whatever their purpose, must be established and operated in such a manner as not to cause harmful interference to the radio services or communications of other Members...', thus clearly including even satellite systems whose purpose is not to be a communication network, but could be data centres in space, or Wireless Power Transport (WPT) satellites, or even space telescopes. Note the preamble uses the term 'all stations' without qualification, and is therefore not limiting the obligation to avoid harmful interference created by the transmission mechanism, but rather the station as a whole. In addition to these general obligations placed on stations, the RRs also further specify requirements on the equipment comprising a station. In RR 3.1: 'The choice and performance of equipment to be used in a station and any emissions therefrom shall satisfy the provisions of these Regulations.' While in this sentence, specific reference is made to *emission*, we consider *Station* as defined above to comprise the necessary accessory equipment and therefore all electromagnetic radiation from the station (and equipment) should be considered, regardless of source.

Moreover, the RRs impose explicit procedural and operational requirements to ensure that harmful interference is avoided. RR 4.3 requires that 'Any new assignment or any change of frequency or other basic characteristic of an existing assignment shall be made in such a way as to avoid causing harmful interference...'. In addition, RR 15.2 and 15.11 necessitate that 'Transmitting stations shall radiate only as much power as is necessary to ensure a satisfactory service.' and 'If, while complying with the provisions of RR 3, a station causes harmful interference [...], special measures shall be taken to eliminate such interference', respectively. It is our view that radiation that is not required for the intended emissions is in excess of the necessary power to ensure the service. RR 3 provides guidance on 'Technical characteristics of stations', with RR 3.3, and 3.6–3.8 providing clear guidance on the obligations to prevent unwanted emissions, and RR 3.11 specifically requires that '...due regard [is] being paid to the Doppler effect where appropriate'.

Taken together, these provisions apply fully to the protection of the RAS: RR 4.6 states that 'the radio astronomy service shall be treated as a radiocommunication service' for interference protection purposes.

In addition to the articles in the RRs, multiple footnotes throughout RR 5 further supplement protections for passive services from interference. For example, 5.149 (administrations are urged to take all practicable steps to protect the RAS from harmful interference) and 5.208B, which defers to ITU-R Res. 739, where recognizing (b) seems to anticipate UEMR 'that, although some unwanted emissions from transmitters on space stations can be controlled through careful design methods and appropriate testing procedures, other unwanted emissions, such as narrowband spurious emissions, generated by uncontrollable and/or unpredictable physical mechanisms, may only be detected after the spacecraft is launched'; the latter referencing radiation in a circuitous way.

These provisions demonstrate that the RRs already contain the principles upon which any type of radiation from satellites can be addressed. What is presently lacking is the operational machinery to give those principles effect: a definition specific to radiation from satellite platform electronics that distinguishes it from the existing categories of unwanted emission; an agreed measurement methodology applicable to on-orbit, fully deployed spacecraft rather than to ground-based anechoic-chamber tests; quantitative compliance criteria against which a satellite or constellation can be assessed; and a pathway for enforcement, noting that the ITU framework relies on coordination between administrations rather than on direct enforcement, and that operations conducted under RR 4.4 sit largely outside the protections afforded by the rest of the RRs. We acknowledge that alternative interpretations of the existing provisions are under active discussion within the relevant ITU-R SGs, and that the view set out above is one of several legitimate readings of the current text. We none the less consider that progress on each of the four operational gaps identified, regardless of which interpretation ultimately prevails, is necessary if the protections envisaged by Preamble 0.4, RR 4.6, and Resolution 739 (Rev. WRC-19) are to be realized in practice for satellite constellations of the scale now being deployed.

### 2.4 Electromagnetic interference standards

Unlike for terrestrial systems, in space applications, there are no mandatory EMI standards. Electromagnetic interference (EMI) standards for terrestrial equipment have thus far adequately addressed interference concerns for radio astronomy.

The exposure of radio astronomy stations to terrestrial EMI sources, even where those sources are non-compliant with applicable EMI standards, is in practice substantially reduced by terrain shielding and clutter loss between the emitter and the telescope. Most radio astronomy observatories are deliberately sited to maximize these effects: at the Paul Wild Observatory the surrounding terrain provides tens of decibels of attenuation against ground-based emitters at the relevant frequencies, and analogous siting decisions apply at Mopra, the Murchison Radio-astronomy Observatory, and other major facilities. Satellite emitters in low-Earth orbit enjoy no such attenuation: the radio path between satellite and telescope is line of sight for the entire portion of the satellite's pass above the local horizon, with only free-space loss and atmospheric absorption between the source and the antenna. The interference problem introduced by large satellite constellations is therefore not simply a scaled-up version of the terrestrial EMI problem; it is qualitatively different, because the propagation mechanisms that protect radio astronomy stations from ground-based emitters do not apply to emitters in orbit.

The satellite industry, however, does apply some EMI/EMC standards in the development of its systems. However, these EMC standards are designed to ensure compatibility within the satellite system itself (i.e. self-interference), and are not sufficient to protect radio astronomy: MIL-STD-461(Department of Defence 2015), ECSS-E-ST-20-07C (European Cooperation for Space Standardization (ECSS) 2022), and NASA Standards (SSP Series–for ISS equipment). Developing appropriate EMI requirements for space systems that protect radio astronomy will require engagement with the relevant international standards bodies, principally the International Special Committee on Radio Interference, which develops civilian electromagnetic compatibility emission limits adopted in many national regulatory frameworks, and the International Organization for Standardization, whose Technical

Committee 20 (Aircraft and Space Vehicles) and its subcommittees develop technical standards for space systems. Neither body currently maintains provisions that specifically address protection of the RAS from radiation originating on operational satellite platforms.

## 3 CURRENT SATELLITE CONSTELLATIONS

As of Feb 2026, the largest constellation is SpaceX's Starlink constellation, providing both internet connectivity from 10.7 to 12.7 GHz (downlink) and 14 to 14.5 GHz (uplink), and DTC services. DTC in some regions is known as Direct-to-Handset, or Direct-to-Device, or Direct-to-Mobile. Going forward, we will refer to this type of service as DTC. DTC allows commercial off-the-shelf consumer mobile phone handsets and tablets (known as User Equipment, or UEs) to connect directly to the satellites – which function in essence as International Mobile Telephony (IMT) base stations in orbit. If existing UEs are being used, that means the bands they use must be existing IMT allocated bands – i.e. bands the mobile devices have transmitters and receivers for. IMT allocations are all terrestrial allocations, the RRs do not yet have any allocated bands for the use of IMT spectrum from space, use of those bands for communication between UEs and satellites is approved by national regulators under ITU-R Article 4.4. Under RR 4.4, national administrations are permitted to allow use of the radio spectrum in derogation of the RRs, as long as they do not claim protection from or cause interference to services allocated in the RRs in other administrations. Considering the long-term unsuitability of RR 4.4 as a vehicle for at-scale IMT-from-space deployments, two Resolutions were approved at the World Radio Conference 2023 (WRC23) that deal with additional allocations, proposing to add new mobile satellite spectrum (MSS) to the table of frequency allocations, some of which was previously IMT terrestrial spectrum: Resolution 252 (WRC23) (ITU-R 2023a), providing low-data rate NGSO MSS, and Resolution 253 (WRC23) (ITU-R 2023b), providing DTC service in MSS spectrum. As a result of this, two Agenda items 1.12 and 1.13, respectively, are being deliberated in the current WRC27 cycle and will be decided at the WRC 2027. Agenda item 1.12 deals with

> ...to consider, based on the results of studies, possible allocations to the mobile satellite service and possible regulatory actions in the frequency bands 1 427-1 432 MHz (space-to-Earth), 1 645.5-1 646.5 MHz (space-to-Earth) (Earth-to-space), 1 880-1 920 MHz (space-to-Earth) (Earth-to-space) and 2 010-2 025 MHz (space-to-Earth) (Earth-to-space) required for the future development of low-data-rate non-geostationary mobile satellite systems, in accordance with Resolution 252 (WRC 23);

and agenda item 1.13 deals with

> ...to consider studies on possible new allocations to the mobile-satellite service for direct connectivity between space stations and International Mobile Telecommunications (IMT) user equipment to complement terrestrial IMT network coverage, in accordance with Resolution 253 (WRC-23)

The astute astronomer reading this will immediately notice that one of the bands under deliberation is directly adjacent to the RAS primary band 1400–1427 MHz. This, along with other unwanted emission concerns (e.g. harmonics) are actively being deliberated, with radio astronomers engaged in spectrum management providing essential input for the studies being done on these topics.

### 3.1 Satellite systems

The following NGSO systems were considered in this paper:

(i) SpaceX's Starlink,[2] a large low-Earth orbit (LEO) internet and DTC communications constellation;
(ii) Eutelsat OneWeb[3] (hereafter OneWeb), another LEO broadband network operating mainly in Ku band with gateways in K$\alpha$ band;
(iii) Amazon Leo[4] (formerly Amazon Kuiper), a satellite internet constellation under development operating in the K$\alpha$ band; and
(iv) GuoWang: Hulianwang-Digui, a Chinese LEO internet constellation (G. Krebs 2026).

Among these, SpaceX's Starlink is notable due to its extensive engagement with the radio astronomy community. SpaceX have been exemplary in collaborating with radio astronomy operators, voluntarily developing mitigation measures such as boresight avoidance (B. D. Nhan et al. 2024) for intended emissions, and, after having been notified of the extensive unwanted emissions and unintended radiation, have started undertaking comprehensive pre-launch characterization of their satellite unwanted emissions and unintended radiation in anechoic chambers. However, ground testing cannot fully reproduce the operational environment in space, the thermal conditions, or the electromagnetic behaviour of fully deployed solar arrays, and as a result, some unintended radiation was identified on-orbit in addition to those measured on the ground. SpaceX are continuing to collaborate closely with CSIRO, the National Radio Astronomy Observatory, and other radio astronomy operators to further mitigate these unwanted emissions and unintended radiation.

### 3.2 Observing methods

Because observations for this project are ongoing as large satellite constellations continue to expand, we report the measurements in the present tense where they describe continuing procedures, and in the past tense for completed observations. The Mopra telescope is used for the majority of the observations. The Mopra telescope is a 22 m diameter radio telescope with a shaped Cassegrain main reflector and shaped subreflector located near the township of Coonabarabran in New South Wales, Australia. It is capable of observing at $L$, $S$, $C$, $X$, and $K$ bands using the very long baseline interferometry (VLBI) receiver and frequency conversion system, and from 30 to 50 GHz and 76 to 116 GHz using the millimetre-wave receiver system.

The VLBI backend has a maximum bandwidth of 64 MHz and is used to step through each of the frequencies with a small 6 MHz overlap to avoid missing detections near the band edges. Covering all bands requires 89 individual observations.

The VLBI backend records the raw voltage data containing two polarizations. The data rate in these files is about 3.6 GB min$^{-1}$. They are post-processed to create power spectra with 1 kHz spectral resolution, providing a Doppler tracking resolution of 273 m s$^{-1}$ at the lowest frequency we use of 1.1 GHz and 37 m s$^{-1}$ at 8.9 GHz, respectively. Post-processing of the raw voltage data into power spectra reduces the file size to 15.4 MB min$^{-1}$. Many of the Mopra observations are uncalibrated and given in 'dB count' only. We consider this sufficient for detection experiments with short 2s integration times – any detected signal above the noise floor is in excess of the telescope's SEFD of 200–500 Jy, depending on band (see Table 2).

[2]https://starlink.com
[3]https://www.eutelsat.com/satellite-network/oneweb-leo-constellation
[4]https://leo.amazon.com

**Table 1.** Satellite Constellation Observation Summary, from tracked observations with Mopra. Active fleet and deorbited satellites as per CelesTrak.org TLEs on 31/10/2025.

| Constellation | Active fleet | Observed satellites | Coverage (per cent) | Deorbited Since obs. | Duration (h) | Number of observations Total | $L$ | $S$ | $C$ | $X$ | $K$ |
|---|---|---|---|---|---|---|---|---|---|---|---|
| Starlink | 8990 | 2156 | 24 | 36 | 260.1 | 3126 | 490 | 945 | 1120 | 233 | 350 |
| Kuiper | 153 | 125 | 81.7 | 2 | 28.6 | 338 | 64 | 61 | 105 | 31 | 79 |
| OneWeb | 651 | 447 | 68.7 | 1 | 62.1 | 762 | 111 | 120 | 131 | 40 | 361 |
| Hulianwang-Digui | 103 | 85 | 82.5 | 1 | 25 | 378 | 52 | 62 | 108 | 51 | 108 |
| Total | 9897 | 2813 | 28.4 | 40 | 375.9 | 4629 | 717 | 1188 | 1464 | 355 | 898 |

**Table 2.** Mopra 22 m detection limits per band ($\Delta\nu = 1$ kHz, $\tau = 2$ s, dual pol., $3\sigma$).

| Band | $\nu$ (MHz) | $T_{sys}$ (K) | $\eta$ | SEFD (Jy) | e.i.r.p. limit (dB($W$/Hz)) 850 km | 1200 km | 2400 km |
|---|---|---|---|---|---|---|---|
| $L$ | 1400 | 31 | 0.69 | 332 | −118.4 | −115.5 | −109.4 |
| $S$ | 2400 | 24 | 0.52 | 332 | −118.4 | −115.5 | −109.4 |
| $C$ | 4400 | 19 | 0.65 | 208 | −120.5 | −117.5 | −111.5 |
| $X$ | 7000 | 22 | 0.67 | 234 | −120.0 | −117.0 | −111.0 |
| $K$ | 20000 | 35 | 0.50 | 505 | −116.6 | −113.6 | −107.6 |

In addition to Mopra, follow-up observations were made with the ATCA, located near Narrabri in New South Wales. ATCA is a six-antenna interferometer also operating across the centimetre and millimetre bands. For this study, a single dish of the ATCA interferometer was used. ATCA's newly commissioned BIGCAT backend (Broadband Integrated GPU Correlator for ATCA) (CSIRO Space & Astronomy 2025) provides a substantial increase in capability for this type of work and was used for follow-up observations of selected detections described in this paper, specifically unwanted emission and intended DTC emission. BIGCAT replaces the previous Compact Array Broadband Backend system with a hybrid FPGA + GPU architecture and increases ATCA's instantaneous bandwidth from 4 to 8 GHz in bands above the 16-cm band (with 2 GHz available in the 16-cm band). It offers a minimum spectral resolution of 0.6 kHz, sufficient to resolve Doppler signatures from NGSO satellites across the full range of orbital velocities sampled here. Shared-risk observations began in October 2025, and the system is now operating in its full 4× 2 GHz configuration. For the purposes of this survey, BIGCAT's high-spectral-resolution mode enables calibrated, high-dynamic-range measurements that are not achievable with Mopra's 64 MHz VLBI backend.

### 3.3 Tracking observations

Table 1 shows to date, a total of 4629 tracking observations were executed since April 2024, amounting to about 375.9 h of telescope time. The tracking observations are planned by obtaining the satellite General Perturbation elements (GPs) in Orbit Mean-Elements Message (OMM) format from the US Strategic Space Command's Spacetrack website (Space-Track 2025). Some readers may be more familiar with the superseded Two Line Element (TLE) format originating in the 1970s, which can no longer accommodate the satellite numbers as they are now too numerous and the NORAD identifiers take up too many characters. GPs is the generic term that describes both TLEs and OMMs. These are used to calculate positions by a suite of PYTHON scripts we developed to filter and create suitable satellite track files. These track files are then used as input to a customized version of the telescope drive software. The Mopra and ATCA telescope drive systems are capable of tracking at a maximum rate of 0.33 deg s$^{-1}$ in elevation and 0.66 deg s$^{-1}$ in azimuth, thus limiting observations of tracked satellite passes in dependence of their orbital altitudes – lower orbits having a higher angular velocity on the sky than higher orbits. For Starlink, a maximum culmination elevation of about 35 deg is trackable. For OneWeb, this increases to about 50 deg elevation due to their higher orbital altitudes.

Satellites were further filtered to only include passes that culminate at higher than 22 deg elevation. This results in a slant range distance to the satellites between 850 km (at culmination) and 1200 km (at rise/set) for the lowest orbits of the Starlink DTC satellites, and up to between 1200 and 2400 km for the OneWeb satellites. Free-space path loss is given by $L_{fs} = 20\log_{10}(4\pi d/\lambda) = 20\log_{10}(d) + 20\log_{10}(f) + \text{const}$, and is therefore frequency-dependent: at fixed range, the loss increases by 6 dB for each doubling of frequency. The effective sensitivity of the system at a given satellite range is, however, approximately band-independent because the gain of a fixed-aperture antenna scales as $G \propto f^2$, partially compensating the $f^2$ dependence of the path loss. Combining the two, the free-space path loss between the satellite and the Mopra aperture varies by less than ± 0.5 dB across the receiver bands used here at fixed range, ranging from approximately 165 dB at 850 km to 175 dB at 2400 km.

The theoretical detection threshold is determined by the system equivalent flux density (SEFD) of the Mopra telescope at each observing frequency, given by SEFD $= 2k_B T_{sys}/A_{eff}$, where $A_{eff} = \eta A_{geom}$ is the effective collecting area. Nominal SEFD values for Mopra are taken from the LBA calibration notes.[5] Using a spectral resolution of $\Delta\nu = 1$ kHz, an integration time of $\tau = 2$ s, dual polarization ($n_{pol} = 2$), and a $3\sigma$ detection threshold, the minimum detectable effective isotropic radiated power (e.i.r.p) spectral density from a satellite ranges from approximately −120 dB($W$/Hz) at 850 km to −108 dB($W$/Hz) at 2400 km. The per-band detection limits are summarized in Table 2.

Because of the limited instantaneous bandwidth available, each 64 MHz wide observation was taken on a different satellite for the SpaceX and OneWeb constellations and repeated multiple times to obtain a statistically significant sample. Some care is taken to repeat the observations for each known satellite version and generation to obtain a better sampling of similar

[5] https://www.atnf.csiro.au/vlbi/dokuwiki/doku.php/lbaops/lbacalibrationnotes/nominalsefd

**Table 3.** Receiver bands and frequency ranges used in this study. All observations used the Mopra 22 m telescope except where noted. The ATCA single-dish follow-up used the BIGCAT backend.

| Band | Frequency range | Telescope |
|---|---|---|
| *L* | 1142–1831 MHz | Mopra |
| *S* | 2192–2830 MHz | Mopra |
| *C* | 4390–6832 MHz | Mopra |
| *X* | 7000–8978 MHz | Mopra |
| *K* | 16–26 GHz | Mopra |
| Follow-up: 1–25 GHz | | ATCA (single dish) |

*Note.* Receiver sensitivity does not drop sharply at band edges; high-power signals may be detected in the band roll-off region beyond the ranges listed here.

satellites. The exact frequency ranges observed are summarized in Table 3. This stepped-band strategy, in which each 64 MHz subband is observed on a different satellite, introduces two potential biases that warrant comment. First, because no single satellite is observed across the full 1–26 GHz range simultaneously, a frequency-dependent emission profile cannot be reconstructed for any individual spacecraft from the Mopra data alone; the per-frequency detection rates reported in Tables A1 – A6 are aggregate statistics across the satellites of a given version observed at that frequency, not the emission spectrum of a representative individual. Secondly, because observations of a given frequency are spread across the campaign rather than concentrated in a single epoch, detection rates at different frequencies sample different mixtures of operational states, firmware revisions, and on-orbit ages of the underlying satellite population; this is in addition to the campaign-wide aggregation effect already discussed in Section 4. Both biases are mitigated, but not eliminated, by reporting detections separately for each known satellite version and generation, so that gross hardware-level changes in the emission profile are visible as differences between tables rather than averaged over within a single table. They are mitigated further by the targeted, calibrated ATCA/BIGCAT follow-up observations described above, which sample multiple frequencies on the same satellite simultaneously and so provide a per-spacecraft consistency check on a subset of the Mopra detections.

### 3.4 Detection and attribution uncertainty

We use the calculated Doppler shift as the metric for attribution of a radiation or emission. We consider this method robust in terms of definitive attribution of a detected radiation originating from an object travelling on the calculated Doppler track. Nevertheless, there are some edge cases worth considering:

(i) Some satellites may have a 'shadow' satellite flying in close proximity. The publicly available GPs do not include any classified satellites for example, so it is possible that in certain cases, we may in fact be attributing the radiation or emission to the wrong satellite. It's not readily possible to put an estimate on that error number, other than it is likely a small number. We have observed unintended radiation in tracked passes that did not match any of the satellites we had GPs for, but they matched frequencies we have detected on Starlink satellites. It stands to reason that these would likely have originated on Starshield satellites, the Defence equivalent to Starlink, which likely shares much of the satellite bus architecture, but for which we do not have any GPs.

(ii) DTC satellites need to Doppler correct their intended emissions for UEs to be able to connect. For this reason, we can't tell which DTC satellite is transmitting a specific emission. For the same reason, we may not be detecting all unwanted emission if it is also Doppler corrected. As a consequence, we would underestimate the number of unintended emissions from these satellites.

(iii) It is conceivable that a terrestrial transmitter could emulate a time variable frequency shift that emulates that of an expected satellite pass. Considering we have observed the same Doppler-shifted radiations with telescopes that are hundreds of kilometres apart simultaneously, we consider this scenario to have an exceedingly remote likelihood.

(iv) Reflections off a satellite from ground-based transmitters would exhibit twice the Doppler shift in frequency and would likely only show up as short blips as the satellite – telescope geometry moves through specular reflective surfaces. We are discounting this based on both accounts.

(v) Reflections between satellites would also result in a combined differential velocity Doppler profile and would not match the GP calculated Doppler profile. For completeness sake, and to satisfy our most ardent critics, we calculated the estimated probability of detecting a reflection of an NGSO satellite's emission or radiation from the NGSO satellite we track for 3 min as negligible, and for an NGSO satellite reflecting a GSO beam into the telescope for a 3 min track as less than 1/10 000 – noting the specular reflection issue mentioned above applies here as well.

## 4 RESULTS

An overview of the distribution of detections across Mopra's observing bands is summarized in Table 4. Detections were categorized by visual inspection with initial numerical sampling as weak (3-5$\sigma$), strong ($> 5\sigma$), or intermittent (where the radiation is observed to disappear and reappear mid-pass) based on their received power relative to the noise floor, the continuity of their Doppler-shifted curves, and their temporal stability across each observation (see the annotations on Fig. 1). The subsequent figures illustrate these detections by band and highlight where detections overlap with radio astronomy allocations (primary, co-primary, secondary, and those protected under RR footnotes such as 5.149 and 5.208B). Further statistics for each unique frequency detection by the operator are presented in the Appendix.

In our measurements, we have detected intended emission, unwanted emission, and unintended radiation from OneWeb satellites at the frequencies detailed in Table A1, Hulianwang-Digui satellites in Table A6, and SpaceX's Starlink satellites in Tables A2 and A3. No detections attributable to Amazon Leo satellites were identified in the observations to date.

The counts presented in Table 4 and Figs 2, 3, 5, 6, and 8 represent the cumulative total of all detections recorded across all satellite systems and across the full observation campaign. They should not be interpreted as representing the current state of any individual satellite system's emission profile. Aggregation suppresses temporal variability across the observing campaign. However, given the evolution of satellite systems a time analysis is required. The current data set does not yet provide a statistically robust sample of individual satellite units. Subsequent papers will present a time analysis of detections including temporal trends and mitigation effectiveness. Several factors complicate a straightforward reading of raw detection counts. First, the

**Table 4.** Satellite detection summary from tracked observations with Mopra.

| Constellation | Number of detections by band (number of unique frequencies) $L$ | $S$ | $C$ | $X$ | $K$ | Total Detections | Total unique Frequencies | Weak | Strong | Interm. | Table |
|---|---|---|---|---|---|---|---|---|---|---|---|
| OneWeb | 6 (6) | 27 (3) | – | 24 (1) | 22 (9) | 79 | 19 | 27 | 31 | 21 | A1 |
| Original Starlink | – | – | 9 (2) | – | – | 9 | 2 | 4 | 2 | 3 | A2 |
| Starlink V2-Mini DTC | 19 (3) | 1762 (111) | 6 (3) | – | – | 1787 | 117 | 199 | 127 | 1461 | A4 |
| Starlink V2-Mini with Ku | 97 (6) | 83 (8) | 38 (10) | 17 (1) | – | 235 | 25 | 75 | 91 | 69 | A3 |
| Starlink V2-Mini | 22 (5) | 18 (5) | – | – | – | 40 | 10 | 7 | 22 | 11 | A5 |
| Hulianwang-Digui | 65 (61) | 137 (79) | 2 (2) | – | – | 204 | 142 | 60 | 108 | 36 | 108 |
| Total | 209 (81) | 2027 (206) | 46 (15) | 41 (2) | 22 (9) | 2345 | 305 | 368 | 379 | 1598 | |

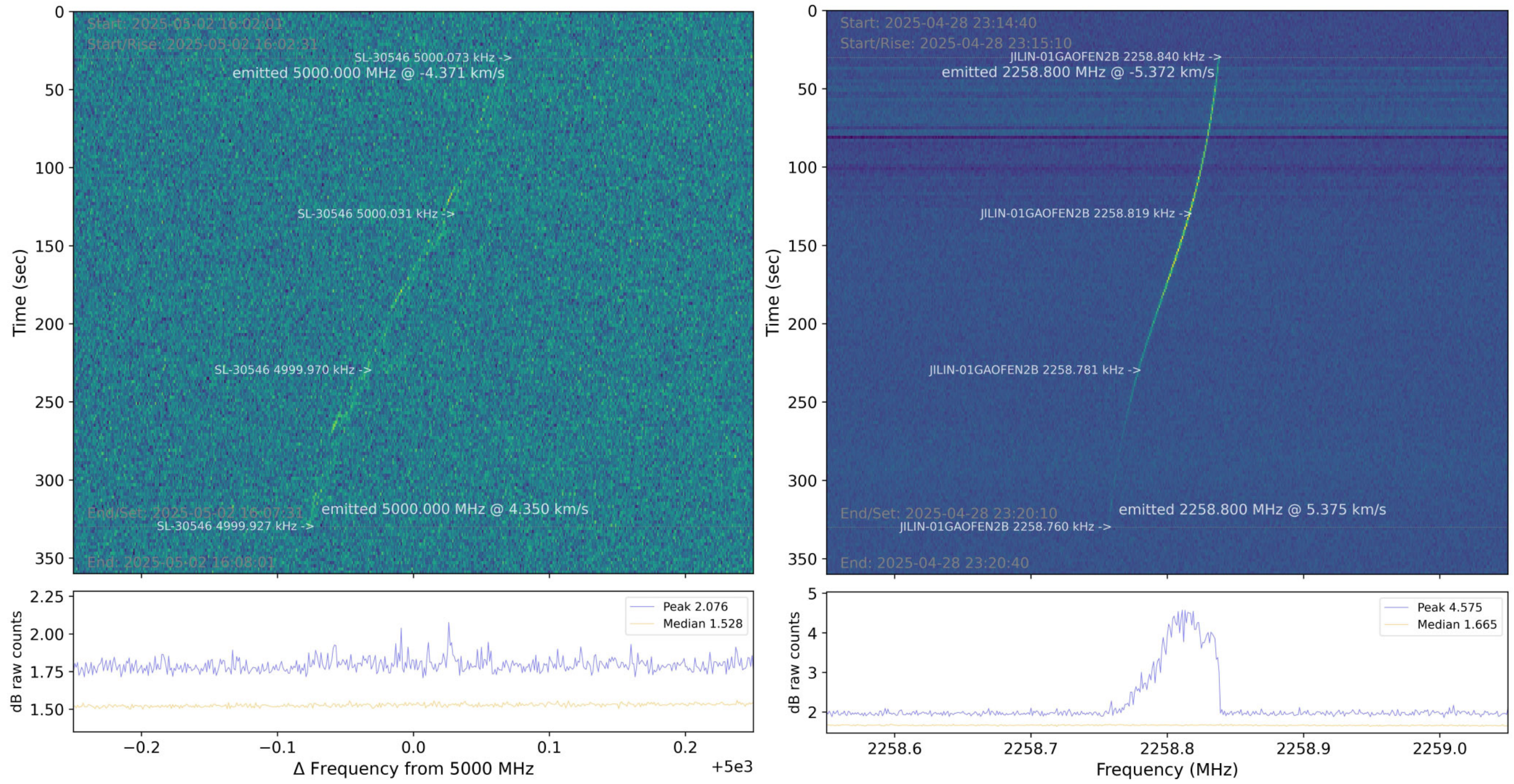

**Figure 1.** Examples of waterfall plots showing the received power for Doppler-shifted curves across Mopra's different frequency bands from various satellite constellations. White vertical lines are masking known local radio-frequency interference or computational artefacts. Detections were categorized as being weak, strong, or intermittent based on their received power relative to the noise floor, the continuity of the curve, and the temporal stability across the observation.

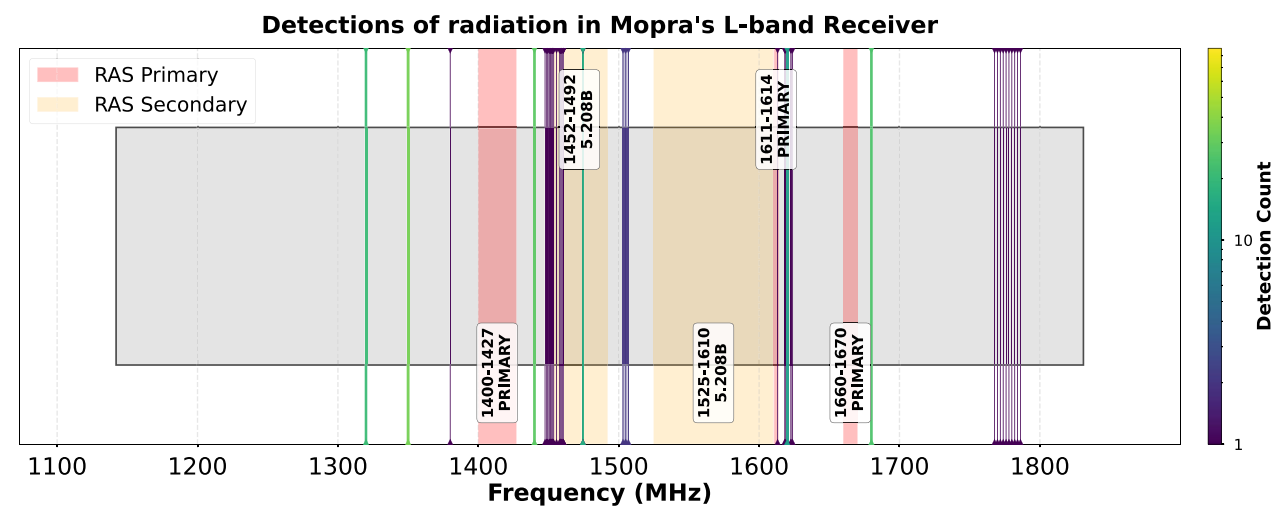

**Figure 2.** $L$-band Mopra detections (1142–1831 MHz). The grey bar shows the receiver band, vertical lines indicate unique frequencies detected, and shaded regions indicate radio astronomy allocations (primary in red and secondary in orange). 81 unique frequencies were detected, including one at 1613.19 MHz within the primary radio astronomy allocation protecting the 1612.231 MHz OH line.

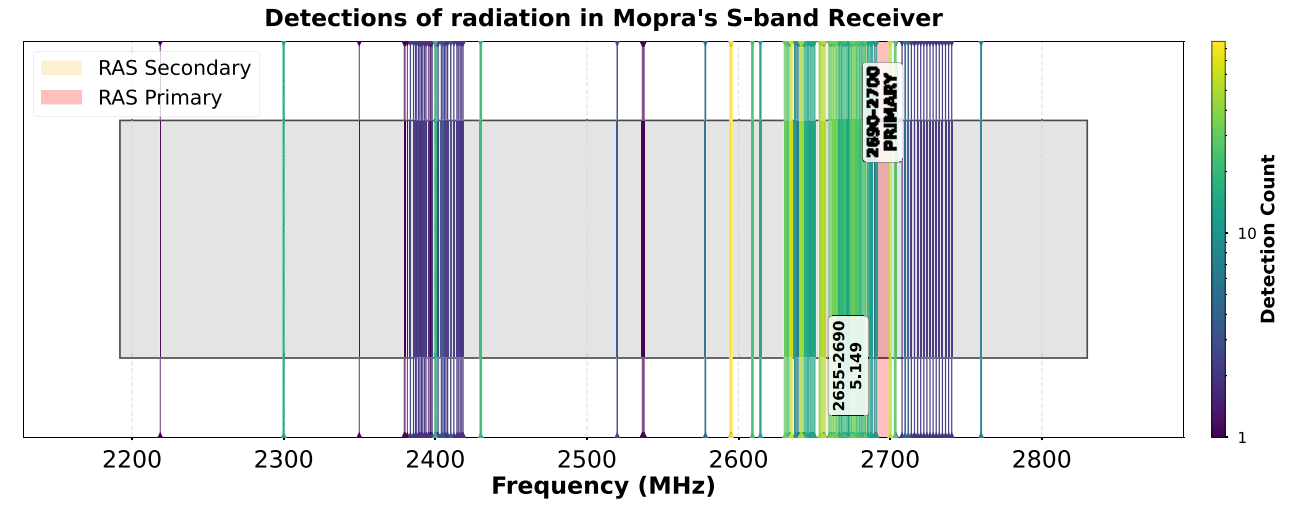

**Figure 3.** Mopra $S$ band (2192–2830 MHz). Grey bar shows receiver band; vertical lines indicate frequency of detections; shaded regions indicate radio astronomy allocations (primary in red and secondary in orange). 206 unique frequencies were detected, with two falling inside the primary radio astronomy band and a cluster near the domestic DTC intended emission at 2620–2630 MHz.

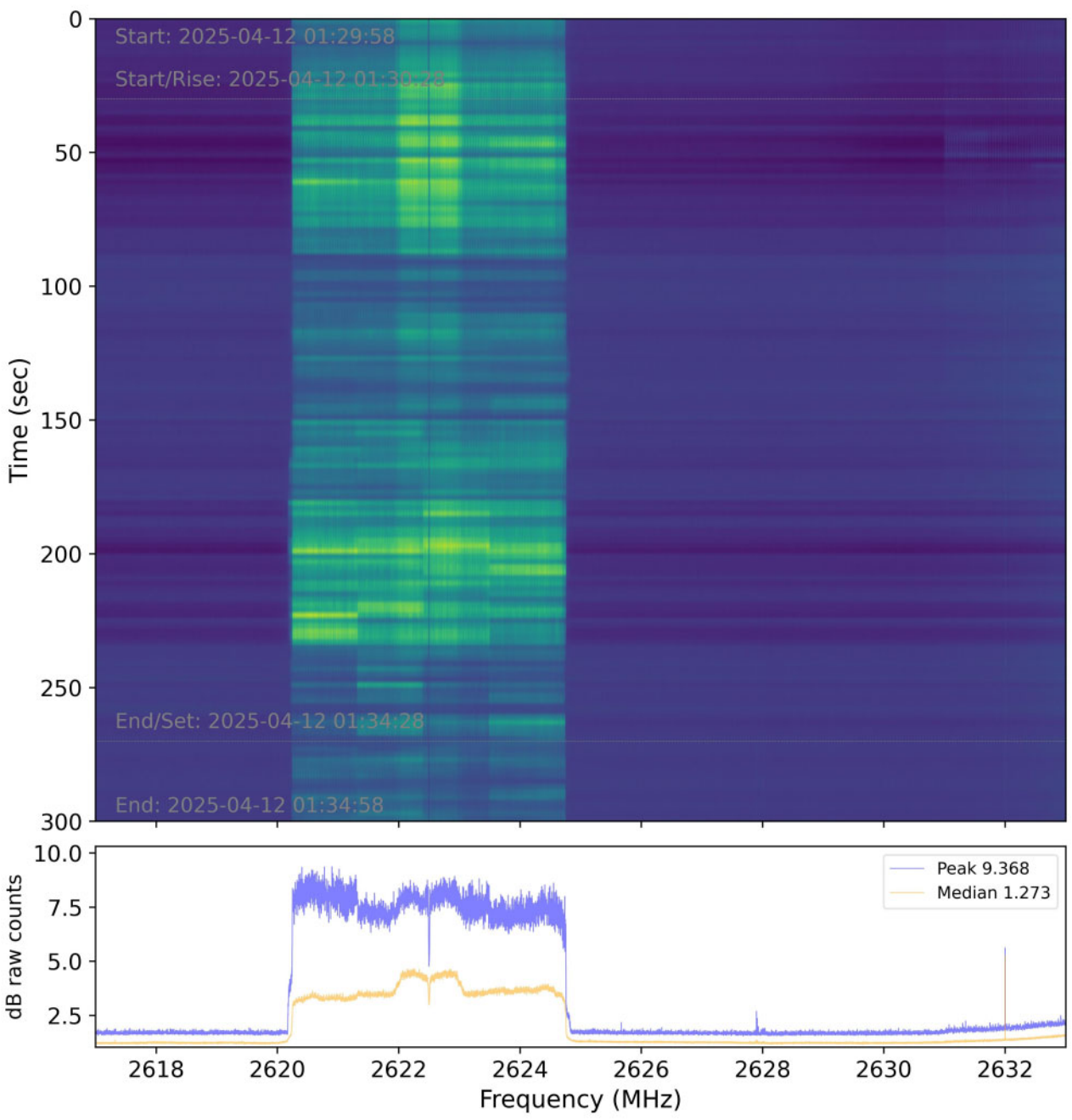

**Figure 4.** Waterfall plot of a Starlink DTC emission, currently the first and only DTC frequency in Australia, showing received power (counts in dB) as a function of frequency (MHz) and time (s). The lower panel displays the median (orange) and max-hold (purple) traces.

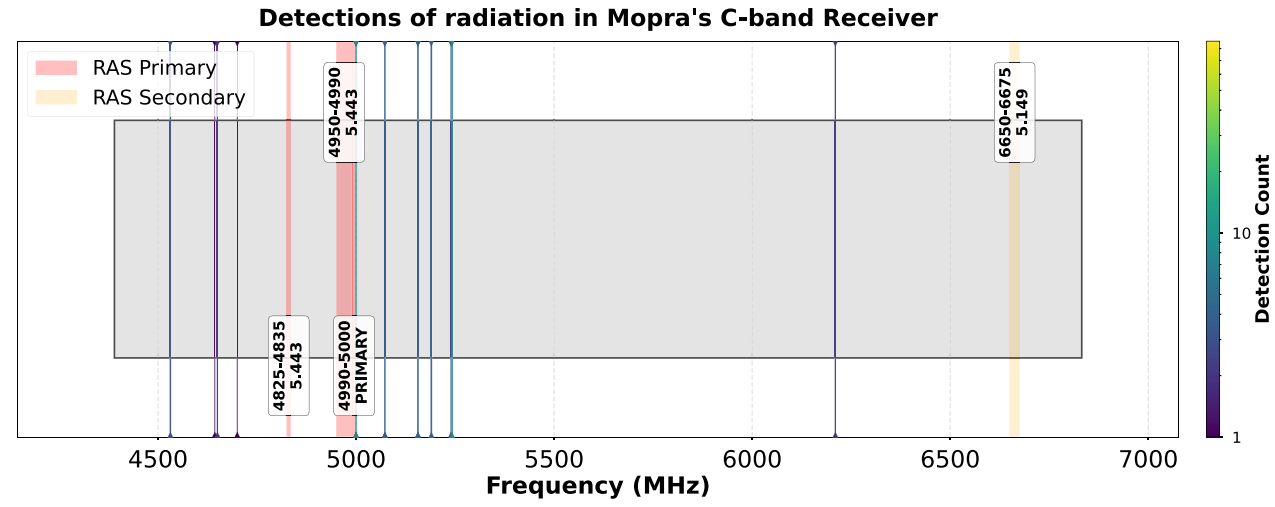

**Figure 5.** *C*-band Mopra detections (4390–6832 MHz). The grey bar shows receiver band; vertical lines indicate frequency of detections; shaded regions indicate radio astronomy allocations (primary in red and secondary in orange). 15 unique frequencies were detected, with one falling inside the primary radio astronomy band.

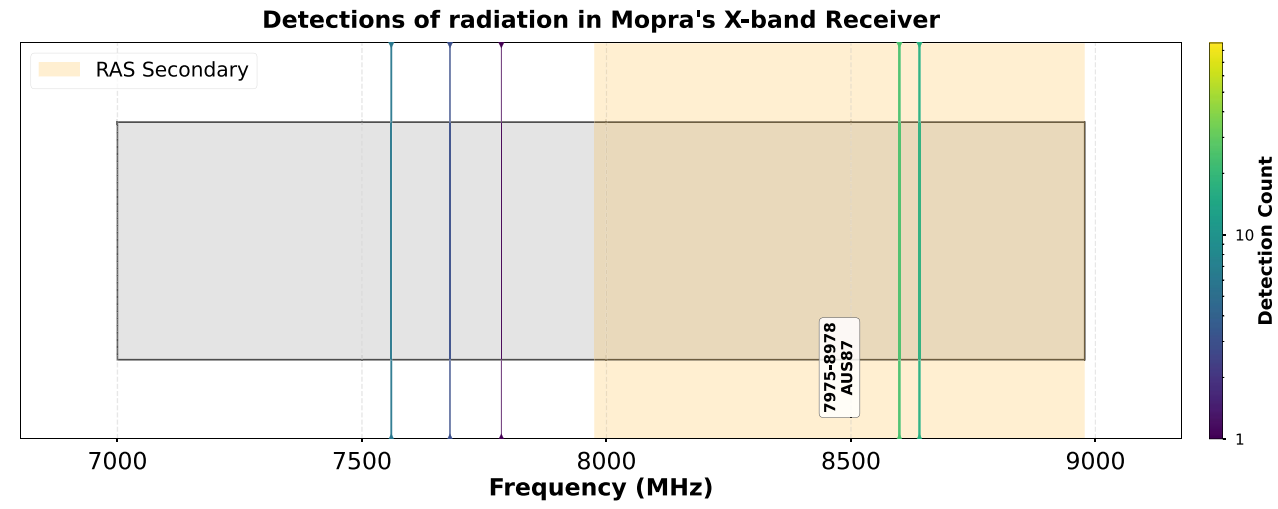

**Figure 6.** *X*-band Mopra detections (7000–8978 MHz). The grey bar shows receiver band; vertical lines indicate the frequency of detections; shaded regions indicate radio astronomy allocations (secondary in orange).

campaign period encompasses active mitigation efforts by satellite operators. The unique detection totals – approaching the maximum of ≈80 detections – therefore represent a historical aggregate rather than a current snapshot. Secondly, in some systems we see variability within the system, where for example one or two satellites are the source of the majority of unintended radiation detections.

## 4.1 *L* band

81 unique frequencies were detected in the operating range of Mopra's *L* band (Fig. 2). The detection density is highest in two clusters: one around 1450 MHz and the other around 1770 MHz. The first includes detections from Hulianwang-Digui and Starlink V2 Mini – DTC satellites (Tables A6 and A4, respectively), some fall within a secondary allocation under Footnote 5.208B. The second cluster originates from a single Hulianwang-Digui satellite.

Notably, 1613.19 MHz falls within the 1610.6–1613.8 MHz primary RAS allocation, which protects radio astronomy observations of the 1612.231 MHz Hydroxyl radical (OH) spectral line, and has been attributed to OneWeb satellites (Table A1). The frequencies detected most often are 1350 , 1440, and 1680 MHz with 34, 28, and 25 instances, respectively.

## 4.2 *S* band

206 unique frequencies were detected in the operating range of Mopra's *S*-band receiver (Fig. 3). The detection density is highest in three clusters: centred about 2400, 2660, and 2725 MHz primarily due to two Hulianwang-Digui satellites, Starlink DTC satellites, and two other Hulianwang-Digui satellites, respectively. Recall, this may not reflect mitigation efforts satellite operators have made since the survey started, rather it represents the aggregate detections over the course of the measurement campaign.

Two detections fall within the primary RAS allocation (Fig. 3): 2690.76 and 2700 MHz. The second detection only affects the protected band when Doppler-shifted as the satellite recedes; this is also the case in Fig. 1(a), where from the centre of the plot to the left, the detection is in a RAS primary band. The second cluster starts just above the Starlink DTC intended emission (shown in Fig. 4) 2620–2630 MHz[6] and extending through the secondary radio astronomy allocation under Footnotes 5.149 and 5.208B. Table A4 shows that most detections in this cluster's Doppler-shifted curves were intermittent in time.

Fig. 4 shows a waterfall plot of early Starlink DTC transmission, SpaceX's Starlink is currently the first and only active DTC transmission provider in Australia operating in the 2620–2630 MHz frequency range. The lower panel displays the median (orange) and max-hold (purple) spectral traces, together revealing both the time-averaged emission profile and the peak power envelope across the observation.

## 4.3 *C* band

It is important to emphasize that this plot shows an intended emission – a nationally licensed DTC downlink operating – rather than unwanted emission or unintended radiation. We include it here to characterize the intended emission, thereby understanding the future risk to radio astronomy, e.g. to illustrate the dynamic-range challenge posed by DTC transmissions, and as a reference against which the harmonics and unintended radiation discussed herein are interpreted. ATCA follow-up measurements show flux densities of order kJy–MJy, many orders of magnitude above typical cosmic sources. The detection of the DTC transmission, at this early stage, provides an opportunity to establish our boresight-avoidance infrastructure. This mitigation technique reduces the harmful effects to the radio astronomy station of beam coupling.

[6]At the time of the measurement in Fig. 4 Starlink DTC was only using a 5 MHz bandwidth.

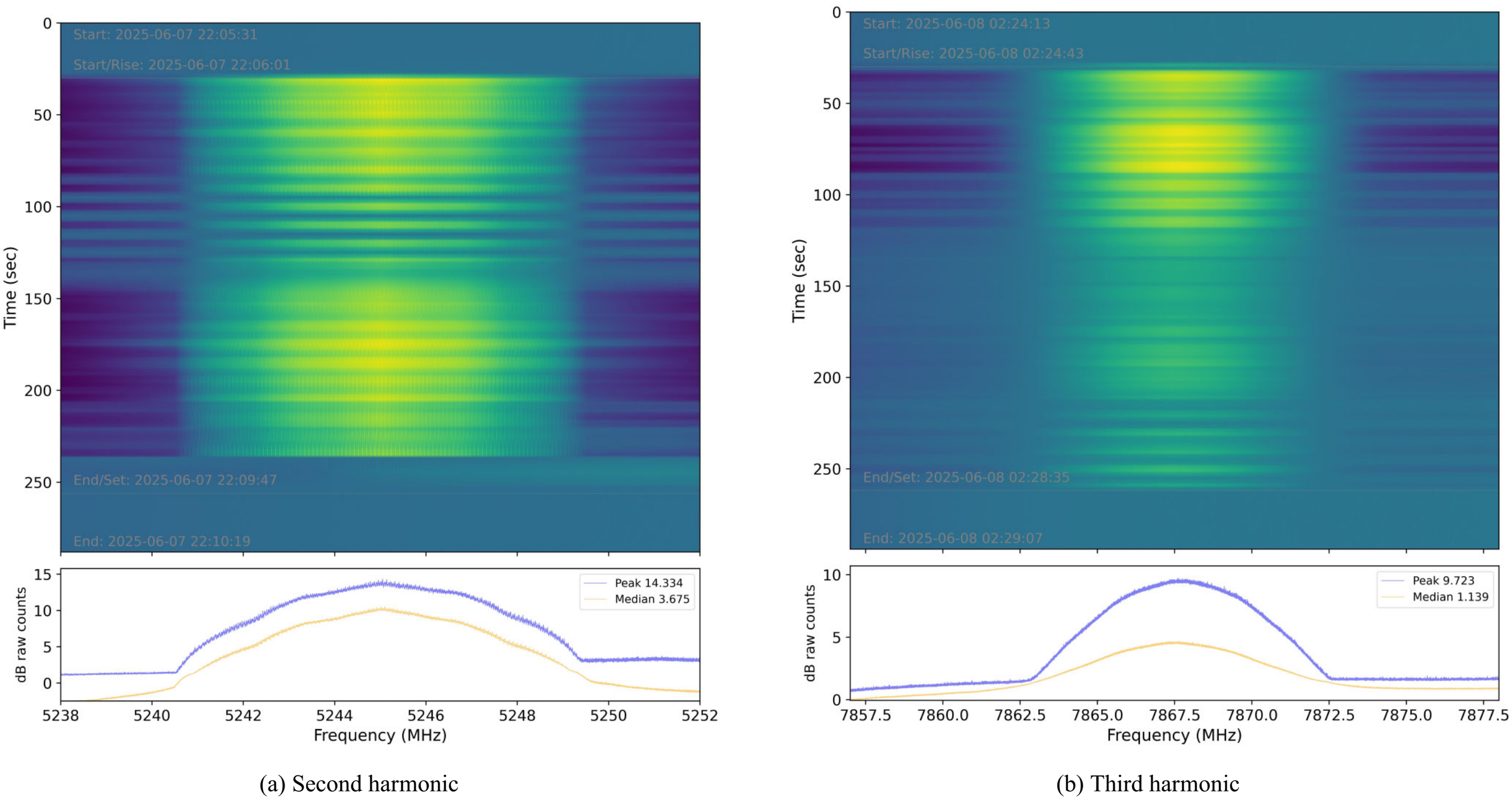

(a) Second harmonic

(b) Third harmonic

**Figure 7.** Waterfall plots showing harmonics of STARLINK DTC signal.

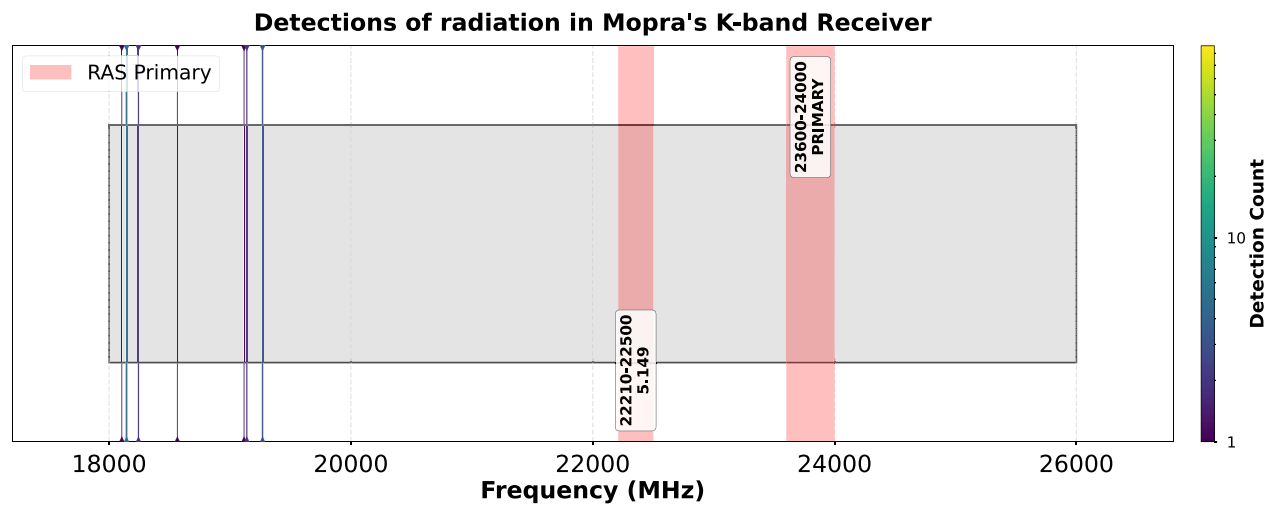

**Figure 8.** *K*-band Mopra detections (18–26 GHz). Grey bar shows receiver band; vertical lines indicate frequency of detections; shaded regions indicate radio astronomy allocations (primary in red).

**Table 5.** Detections of Harmonics, from tracked observations with Mopra.

| Operator | Harmonic | Frequency (MHz) | No. of detections |
|---|---|---|---|
| Starlink (DTC) | Second | 5240–5260 | 7 |
| Starlink (DTC) | Third | 7860–7890 | 2 |
| Starlink (DTC) | Fourth | 10480–10520 | |

15 unique frequencies were detected in Mopra's *C* band. One detection, 5000 MHz, falls in a RAS primary band (Fig. 1a) when the satellite is receding – similar to the 2700 MHz detection in the *S* band, with both associated with Starlink satellites (Tables A3 and A5). 5000 MHz also had the most detections in this band. In addition, the second harmonic of the Starlink DTC signal was identified in the *C* band between 5240 and 5260 MHz (Fig. 7a and Table 5). Note that part of *C* band is covered by Mopra's *X*-band receiver (Fig. 6). Three unique frequencies were detected in this frequency range corresponding to early Starlink satellite units (Table A2).

### 4.4 *X* band

Although Fig. 6 shows five apparent detections, only 2 fall strictly within *X*-band and are thus represented in Table 4 as such. The two detections are associated with OneWeb and Starlink V2 Mini satellites (Tables A1 and A5). Additionally, the third harmonic of the Starlink DTC signal was identified (Fig. 7b and Table 5), spanning 7860–7890 MHz. 8599 MHz is the most frequently detected unintended radiation, corresponding to OneWeb satellites. The fourth harmonic of the Starlink DTC signal 10480–10520 MHz (Table 5) was identified while commissioning the ATCA/BIGCAT backend, but is not presented here.

### 4.5 *K* band

Nine unique frequencies were detected across Mopra's *K* band, all attributed to OneWeb satellites (Table A1). No overlap with radio astronomy allocations was observed in this band. We note that the nine detected frequencies are all in a Fixed Satellite Service (FSS) PRIMARY (S→E) allocation. Since these frequencies are in an FSS primary allocation they may be operational in nature (i.e. intended emission rather than unintended radiation).

## 5 CONCLUSIONS

We have presented the first results of SNIFFLES, an ongoing survey of intended emission, unwanted emission, and unintended radiation from NGSO systems, conducted with the Mopra radio telescope, with targeted follow-up using a single dish of the ATCA interferometer. Since April 2024, 4629 tracked observations – totalling 375.9 h across Mopra's *L*, *S*, *C*, *X*, and *K* receiver bands (1–26 GHz) – yielded 2345 detections at more than 300 unique frequencies. These detections are from OneWeb, original Starlink, Starlink V2-Mini (including DTC-enabled & Ku versions), and Hulianwang–Digui satellite systems. No detections were attributed to Amazon Leo, consistent with it not yet being operational. Several detections fall in primary RAS allocations: 1613.19, 2690.76, 2700, and 5000 MHz, though the latter two are only in the protected band when the satellite is receding. Of particular note, unintended radiation at 2700 MHz was detected in 76.9 per cent of all observations of the relevant Starlink V2-Mini satellite version (Table A3), underscoring the persistence of this radiation within a primary radio astronomy allocation. We note that the Mopra receivers begin at *L* band (1142 MHz); characterization of the 50–350 MHz range covered by other studies is not addressed here.

The observed interference falls into three classes: (1) intended emission, including strong MJy DTC signals which risk saturating receivers during beam-coupling; (2) unwanted emission (out-of-band emission, including harmonics), up to the fourth harmonic of the Starlink DTC transmission; and (3) unintended radiation from satellite platform electronics and subsystems.

Section 2 showed the ITU-R RRs already provide a basis for addressing unwanted emissions and unintended radiation from satellite platforms. Without intervention, continued growth of unintended radiation risks degrades not only radio astronomy (our window into the universe) but also the geodetic and timing infrastructure underpinning Global Navigation Satellite Systems and satellite operations – critical space-infrastructure services on which modern society depends.

We are currently undertaking observations in Mopra's Q- (30–50 GHz) and *W* bands (76–116 GHz). Several satellite systems remain to be characterized, including Lynk Towers 1–6, AST SpaceMobile/Bluewalker 3, Qianfan, and Earth-observation systems (e.g. Fig. 1a) – with the Amazon Leo system requiring follow-up when operational. As these systems mature, mitigation efforts will need to be evaluated, and any new providers or changes to DTC in Australia will require characterization. Future SNIFFLES papers will present calibrated, high-dynamic-range measurements using ATCA/BIGCAT, and probe frequencies below 1 GHz using ASKAP and Murriyang, the 64 m Parkes radio telescope's Ultra-Wideband Low receiver. Although Murriyang cannot track LEO satellites due to its large diameter and slower slew rate, its exceptional sensitivity makes it valuable for static measurements of DTC and for verifying boresight avoidance (Indermuehle 2024). The impact of satellite systems on future ATNF receivers (CSIRO Space & Astronomy 2024) – including the Murriyang cryoPAF, and aperture arrays under development – and the efficacy of boresight avoidance over time will also be investigated.

Through these efforts, SNIFFLES aims to provide the empirical basis for evaluating mitigation, supporting the development of EMI standards for space systems, and to demonstrate that sustainable long-term coexistence of NGSO constellations and ground-based radio astronomy is both necessary and achievable. As the observational data set grows in temporal coverage and regularity, it will enable a time-resolved assessment of detection rates and their dependence on satellite generation.

## ACKNOWLEDGEMENTS

The Mopra and ATCA telescopes are part of the ATNF that is funded by the Australian Government for operation as a National Facility managed by CSIRO, Australia's national science agency. We acknowledge the Gomeroi people as the Traditional Owners of the Paul Wild Observatory site. We are grateful to the referee for a careful and constructive report that substantially improved the clarity of the text and the presentation of our methodology.

## DATA AVAILABILITY

The observation data and the data base of detections presented herein are not publicly available at this time, pending consultation with satellite operators regarding responsible disclosure.

## APPENDIX A: DETAILED TABLES OF FREQUENCIES DETECTED USING MOPRA

This appendix presents the complete tables listing the detected frequencies associated with each satellite system and satellite generation discussed in the main text. The tables provide the centre frequency, number of detections, qualitative detection classification, detection rate calculated from the total observations, and relevant regulatory footnotes. This data complements the aggregate summaries presented in Section 4 and provides a more detailed breakdown of the observed radiation identified throughout this Satellite measuremeNt of Intended emission, unwanted emission, and radio radiation to develop, Follow up, and veriFy mitigation measures, regulatory compliance, and Lawful usE of the radio Spectrum (SNIFFLES) observing campaign.

The detection counts presented here represent cumulative observations across the full survey period and should therefore be interpreted as historical aggregates rather than snapshots of the current behaviour of any individual satellite system.

Table A1 presents detections associated with OneWeb. Tables A2–A5 summarize detections associated with different generations and versions of Starlink, including original Starlink satellites, Starlink V2-Mini Ku-optimized satellites, Starlink V2-Mini satellites equipped with DTC capability, and 'standard' Starlink V2-Mini satellites without DTC capability. Table A6 presents detections attributed to Hulianwang-Digui.

**Table A1.** Detections for OneWeb Satellites, from tracked observations with Mopra. Numbers in brackets indicate unique satellites. All detections in the table are also covered by footnotes AUS87 and L-band detections are covered by footnote AUS103.

| Frequency [MHz] | Total Detections (# unique satellites) | Weak | Strong | Interm. | Detection Rate (per cent) | Total Observations | ARSP 2021 footnotes | Allocation |
|---|---|---|---|---|---|---|---|---|
| 1613.19 | 1 | – | – | 1 | 12.50 | 8 | 5.149 | RAS primary |
| 2218.56 | 1 | – | – | 1 | 16.7 | 6 | | |
| 2300.00 | 14 (12) | 11 | 3 | – | 93.3 | 15 | | |
| 2400.00 | 12 | 7 | 5 | – | 85.7 | 14 | | |
| 8599.00 | 24 | 7 | 16 | 1 | 100 | 24 | | |
| 17837.03 | 4 | – | 1 | 3 | 66.7 | 6 | | FSS primary (s-E) |
| 17876.63 | 3 | – | – | 3 | 50 | 6 | | FSS primary (s-E) |
| 18105.20 | 1 | – | – | 1 | 25 | 4 | | FSS primary (s-E) |
| 18144.14 | 5 | – | – | 5 | 100 | 5 | | FSS primary (s-E) |
| 18242.79 | 2 | – | 2 | - | 66.7 | 3 | | FSS primary (s-E) |
| 18564.43 | 1 | – | – | 1 | 33.3 | 3 | | FSS primary (s-E) |
| 19116.14 | 1 | – | – | 1 | 25 | 4 | | FSS primary (s-E) |
| 19139.04 | 2 | – | 1 | 1 | 50 | 4 | | FSS primary (s-E) |
| 19268.96 | 3 | – | – | 3 | 60 | 5 | | FSS primary (s-E) |

**Table A2.** Detections for original Starlink Satellites, from tracked observations with Mopra. All detections in the table are also covered by footnote AUS87.

| Frequency [MHz] | Total Detections (# unique satellites) | Weak | Strong | Interm. | Detection rate (per cent) | Total observations | ARSP 2021 footnotes | Allocation |
|---|---|---|---|---|---|---|---|---|
| 7560.00 | 6 | 3 | 1 | 2 | 100 | 6 | | FSS primary (s-E) |
| 7680.00 | 3 | 1 | 1 | 1 | 100 | 3 | | FSS primary (s-E) |

**Table A3.** Detections for Starlink V2 Mini - Ku Optimized Satellites, from tracked observations with Mopra.

| Frequency [MHz] | Total Detections (# unique satellites[a]) | Weak | Strong | Interm. | Detection rate (per cent) | Total observations | ARSP 2021 footnotes[b] | Allocation |
|---|---|---|---|---|---|---|---|---|
| 1320.00[c] | 1 | – | – | 1 | 10 | 10 | | |
| 1350.00[c] | 12 | 2 | 8 | 2 | 100 | 12 | | |
| 1620.00 | 7 (6) | 1 | 5 | 1 | 87.5 | 8 | | |
| 1622.85 | 1 | 1 | – | – | 12.5 | 8 | | MSS secondary (s-E) |
| 1680.00[c] | 1 | 1 | – | – | 100 | 1 | | |
| 2430.00[c] | 4 | 1 | 3 | – | 100 | 4 | | |
| 2578.13 | 2 | 1 | – | 1 | 66.7 | 3 | | |
| 2656.44 | 1 | – | – | 1 | 20 | 5 | 5.149, 5.208B | RAS secondary |
| 2700.00[c] | 10 | – | 6 | 4 | 76.9 | 13 | 5.340[d] | RAS primary |

[a]Where the number of unique satellites with detections is less than the total detections.
[b]All detections in the table are also covered by AUS87 and *L*-band detections are additionally covered by AUS103.
[c]Appear in multiple tables, otherwise unique to the version in the table.
[d]The Doppler shift when the satellite is receding results in the radiation being in the protected band.

**Table A4.** Detections for Starlink V2 Mini DTC Satellites, from tracked observations with Mopra.

| Frequency [MHz] | Total detections (unique satellites[a]) | Weak | Strong | Interm. | Detection rate (per cent) | Total observations | ARSP 2021 (footnotes[b]) | ATCA follow-up flux density (Jy) Mean | ATCA follow-up flux density (Jy) Max | Allocation |
|---|---|---|---|---|---|---|---|---|---|---|
| 1440.00[c] | 2 | 2 | – | – | 6.8 | 44 | | | | |
| 1474.50[c] | 15 | 2 | 8 | 5 | 88.2 | 17 | 5.208B | | | |
| 1474.56 | 1 | 1 | – | – | 5.9 | 17 | 5.208B | | | |
| 2595.00 | 86 (83) | 12 | 40 | 34 | 86.9 | 99 | | 1.06kJy – 3.96kJy | 1.89E4 3.47E6 2.13E3 | |
| 2609.20 | 23 (22) | 1 | – | 22 | 11.6 | 198 | | | | |
| 2614.44 | 13 | – | – | 13 | 6.6 | 198 | | | | |
| 2630.76 | 31 (30) | – | – | 31 | 13 | 198 | | | | |
| 2631.24 | 24 | – | – | 24 | 10 | 239 | | | | |
| 2631.96 | 25 | 1 | – | 24 | 10.5 | 239 | | | | |
| 2633.64 | 13 | – | – | 13 | 5.6 | 230 | | | | |
| 2633.88 | 2 | – | – | 2 | 0.9 | 230 | | | | |
| 2634.10 | 18 | 1 | – | 17 | 7.8 | 230 | | | | |
| 2634.60 | 63 (60) | 6 | 1 | 56 | 27.4 | 230 | | | | |
| 2637.24 | 11 | – | – | 11 | 5 | 221 | | | | |
| 2638.68 | 6 | – | – | 6 | 2.7 | 221 | | | | |
| 2639.04 | 4 | – | – | 4 | 1.8 | 221 | | | | |
| 2639.40 | 7 | – | – | 7 | 3.2 | 221 | | | | |
| 2640.12 | 8 | – | – | 8 | 3.6 | 221 | | | | |
| 2640.36 | 5 | – | – | 5 | 2.3 | 221 | | | | |
| 2640.84 | 46 (44) | 4 | – | 42 | 20.8 | 221 | | | | |
| 2641.08 | 23 | 1 | – | 22 | 10.4 | 221 | | | | |
| 2641.80 | 18 | 2 | – | 16 | 8.1 | 221 | | | | |
| 2641.92 | 46 | 6 | – | 40 | 20.8 | 221 | | | | |
| 2642.16 | 9 | – | – | 9 | 6.2 | 144 | | | | |
| 2642.88 | 8 | – | – | 8 | 5.6 | 144 | | | | |
| 2643.12 | 13 | 1 | – | 12 | 9 | 144 | | | | |
| 2643.24 | 5 | – | – | 5 | 3.5 | 144 | | | | |
| 2643.48 | 7 | – | – | 7 | 4.9 | 144 | | | | |
| 2643.72 | 15 | 1 | – | 14 | 10.4 | 144 | | | | |
| 2643.96 | 9 | – | – | 9 | 6.2 | 144 | | | | |
| 2644.44 | 16 | 2 | – | 14 | 11.4 | 141 | | | | |
| 2645.64 | 13 | 1 | – | 12 | 9.2 | 141 | | | | |
| 2647.32 | 5 | 1 | – | 4 | 3.6 | 141 | | | | |
| 2647.42 | 9 | – | 7 | 2 | 6.4 | 141 | | | | |
| 2647.56 | 12 | 1 | – | 11 | 8.5 | 141 | | | | |
| 2648.04 | 6 | – | – | 6 | 3.2 | 189 | | | | |
| 2648.52 | 14 | 1 | – | 13 | 7.4 | 189 | | | | |
| 2648.76 | 9 | 1 | – | 8 | 4.8 | 189 | | | | |

**Table A4** – *continued*

| Frequency [MHz] | Total detections (unique satellites[a]) | Weak | Strong | Interm. | Detection rate (per cent) | Total observations | ARSP 2021 (footnotes[b]) | ATCA follow-up flux density (Jy) Mean | Max | Allocation |
|---|---|---|---|---|---|---|---|---|---|---|
| 2648.88 | 7 | – | – | 7 | 3.7 | 189 | | | | |
| 2649.00 | 8 | – | – | 8 | 4.2 | 189 | | | | |
| 2649.48 | 10 | 1 | – | 9 | 5.3 | 189 | | | | |
| 2649.60 | 8 | – | – | 8 | 4.2 | 189 | | | | |
| 2649.72 | 7 | – | – | 7 | 3.7 | 189 | | | | |
| 2649.96 | 14 | 1 | – | 13 | 7.4 | 189 | | 1.22E3 | 1.86E3 | |
| 2653.80 | 39 (38) | 2 | 1 | 36 | 20.6 | 189 | | | | |
| 2655.00 | 46 (45) | 2 | 32 | 12 | 24.3 | 189 | 5.149, 5.208B | | | RAS secondary |
| 2656.44 | 54 (51) | 16 | 1 | 37 | 28.6 | 189 | 5.149, 5.208B | 4.38E3 | 3.19E4 | RAS secondary |
| 2660.04 | 14 | 1 | – | 13 | 7.4 | 189 | 5.149, 5.208B | | | RAS secondary |
| 2660.28 | 20 | 3 | – | 17 | 10.6 | 189 | 5.149, 5.208B | | | RAS secondary |
| 2660.40 | 15 | 3 | – | 12 | 7.9 | 189 | 5.149, 5.208B | | | RAS secondary |
| 2660.52 | 31 | 8 | – | 23 | 16.4 | 189 | 5.149, 5.208B | | | RAS secondary |
| 2661.00 | 17 | – | – | 17 | 9 | 189 | 5.149, 5.208B | | | RAS secondary |
| 2661.12 | 19 | 1 | – | 18 | 10 | 189 | 5.149, 5.208B | | | RAS secondary |
| 2661.24 | 4 | – | – | 4 | 2.1 | 189 | 5.149, 5.208B | | | RAS secondary |
| 2662.44 | 10 | – | – | 10 | 6 | 167 | 5.149, 5.208B | | | RAS secondary |
| 2662.56 | 2 | – | – | 2 | 1.2 | 167 | 5.149, 5.208B | | | RAS secondary |
| 2662.58 | 34 | 8 | 20 | 6 | 20.4 | 167 | 5.149, 5.208B | | | RAS secondary |
| 2662.68 | 8 | – | – | 8 | 4.8 | 167 | 5.149, 5.208B | | | RAS secondary |
| 2662.92 | 10 | 1 | – | 9 | 6 | 167 | 5.149, 5.208B | | | RAS secondary |
| 2663.16 | 9 | – | – | 9 | 6 | 167 | 5.149, 5.208B | | | RAS secondary |
| 2663.28 | 19 | 3 | – | 16 | 11.4 | 167 | 5.149, 5.208B | | | RAS secondary |
| 2664.36 | 33 | 10 | 1 | 22 | 37.1 | 89 | 5.149, 5.208B | | | RAS secondary |
| 2666.04 | 13 | 1 | – | 12 | 14.6 | 89 | 5.149, 5.208B | | | RAS secondary |
| 2666.28 | 17 | 2 | – | 15 | 19.1 | 89 | 5.149, 5.208B | | | RAS secondary |
| 2666.52 | 17 | 3 | – | 14 | 19.1 | 89 | 5.149, 5.208B | | | RAS secondary |
| 2666.76 | 15 | 1 | – | 14 | 16.8 | 89 | 5.149, 5.208B | | | RAS secondary |
| 2667.96 | 19 | 2 | – | 17 | 21.4 | 89 | 5.149, 5.208B | | | RAS secondary |
| 2668.20 | 13 | – | – | 13 | 14.6 | 89 | 5.149, 5.208B | | | RAS secondary |
| 2669.16 | 12 | 1 | – | 11 | 13.5 | 89 | 5.149, 5.208B | | | RAS secondary |
| 2669.64 | 13 | 1 | 1 | 11 | 14.6 | 89 | 5.149, 5.208B | | | RAS secondary |
| 2669.88 | 17 | 1 | – | 16 | 19.1 | 89 | 5.149, 5.208B | | | RAS secondary |
| 2670.00 | 8 | – | – | 8 | 9 | 89 | 5.149, 5.208B | | | RAS secondary |
| 2670.12 | 3 | – | – | 3 | 3.4 | 89 | 5.149, 5.208B | | | RAS secondary |
| 2670.16 | 22 | 10 | – | 12 | 24.7 | 89 | 5.149, 5.208B | | | RAS secondary |
| 2670.60 | 16 | 1 | – | 15 | 18 | 89 | 5.149, 5.208B | | | RAS secondary |
| 2670.84 | 5 | 1 | – | 4 | 5.6 | 89 | 5.149, 5.208B | | | RAS secondary |
| 2670.96 | 13 | – | – | 13 | 14.6 | 89 | 5.149, 5.208B | | | RAS secondary |
| 2671.20 | 4 | – | – | 4 | 4.5 | 89 | 5.149, 5.208B | | | RAS secondary |
| 2671.32 | 7 | – | – | 7 | 7.9 | 89 | 5.149, 5.208B | | | RAS secondary |

**Table A4** – *continued*

| Frequency [MHz] | Total detections (unique satellites[a]) | Weak | Strong | Interm. | Detection rate (per cent) | Total observations | ARSP 2021 (footnotes[b]) | ATCA follow-up flux density (Jy) Mean | Max | Allocation |
|---|---|---|---|---|---|---|---|---|---|---|
| 2672.52 | 10 | – | – | 10 | 11.2 | 89 | 5.149, 5.208B | | | RAS secondary |
| 2672.76 | 14 | – | – | 14 | 15.7 | 89 | 5.149, 5.208B | | | RAS secondary |
| 2674.20 | 8 | – | – | 8 | 9 | 89 | 5.149, 5.208B | | | RAS secondary |
| 2674.30 | 13 | 5 | 3 | 5 | 14.6 | 89 | 5.149, 5.208B | | | RAS secondary |
| 2674.44 | 29 | 10 | – | 19 | 32.6 | 89 | 5.149, 5.208B | | | RAS secondary |
| 2674.56 | 7 | – | – | 7 | 7.9 | 89 | 5.149, 5.208B | | | RAS secondary |
| 2674.68 | 15 | 3 | – | 12 | 16.8 | 89 | 5.149, 5.208B | | | RAS secondary |
| 2674.80 | 9 | – | – | 9 | 10.1 | 89 | 5.149, 5.208B | | | RAS secondary |
| 2674.92 | 15 | – | – | 15 | 16.8 | 89 | 5.149, 5.208B | | | RAS secondary |
| 2675.76 | 8 | – | – | 8 | 9 | 89 | 5.149, 5.208B | | | RAS secondary |
| 2675.88 | 31 | 8 | – | 23 | 34.8 | 89 | 5.149, 5.208B | | | RAS secondary |
| 2676.36 | 23 | 6 | – | 17 | 25.8 | 89 | 5.149, 5.208B | | | RAS secondary |
| 2676.48 | 15 | 1 | – | 14 | 16.8 | 89 | 5.149, 5.208B | | | RAS secondary |
| 2678.28 | 25 | 5 | – | 20 | 28.1 | 89 | 5.149, 5.208B | 1.93E3 | 4.40E3 | RAS secondary |
| 2678.40 | 25 | 5 | – | 20 | 28.1 | 89 | 5.149, 5.208B | | | RAS secondary |
| 2679.36 | 11 | – | – | 11 | 12.4 | 89 | 5.149, 5.208B | | | RAS secondary |
| 2679.72 | 22 | 5 | – | 17 | 24.7 | 89 | 5.149, 5.208B | | | RAS secondary |
| 2680.92 | 13 | – | – | 13 | 14.6 | 89 | 5.149, 5.208B | | | RAS secondary |
| 2681.16 | 19 | 2 | – | 17 | 21.4 | 89 | 5.149, 5.208B | | | RAS secondary |
| 2681.64 | 16 | – | – | 16 | 18 | 89 | 5.149, 5.208B | | | RAS secondary |
| 2682.60 | 16 | 4 | – | 12 | 18 | 89 | 5.149, 5.208B | | | RAS secondary |
| 2683.44 | 12 | – | – | 12 | 13.5 | 89 | 5.149, 5.208B | | | RAS secondary |
| 2683.56 | 27 | 9 | – | 18 | 30.3 | 89 | 5.149, 5.208B | | | RAS secondary |
| 2683.68 | 2 | – | – | 2 | 2.2 | 89 | 5.149, 5.208B | | | RAS secondary |
| 2685.36 | 4 | – | – | 4 | 4.5 | 89 | 5.149, 5.208B | | | RAS secondary |
| 2685.48 | 14 | – | – | 14 | 15.7 | 89 | 5.149, 5.208B | | | RAS secondary |
| 2687.16 | 13 | – | – | 13 | 4.6 | 89 | 5.149, 5.208B | | | RAS secondary |
| 2687.40 | 8 | – | – | 8 | 9 | 89 | 5.149, 5.208B | | | RAS secondary |
| 2687.88 | 11 | – | – | 11 | 12.4 | 89 | 5.149, 5.208B | | | RAS secondary |
| 2689.80 | 5 | – | – | 5 | 5.6 | 89 | 5.149, 5.208B | | | RAS secondary |
| 2690.28 | 5 | – | – | 5 | 5.6 | 89 | 5.149, 5.208B | | | RAS secondary |
| 2690.76 | 12 | – | – | 12 | 13.5 | 89 | 5.340 | | | RAS primary |
| 2703.36 | 27 | 3 | 10 | 14 | 56.2 | 48 | | 3.48E3 | 2.47E4 | |
| 4643.38 | 1 | – | – | 1 | 8.3 | 12 | | | | FSS primary (s-E) |
| 5190.00 | 4 | – | 2 | 2 | 66.7 | 6 | | | | |
| 7785.00 | 1 | 1 | – | – | 14.3 | 7 | | | | |

[a]Where the number of unique satellites with detections is less than the total detections.
[b]All detections in the table are also covered by AUS87 and *L*-band detections are additionally covered by AUS103.
[c]Appear in multiple tables, otherwise unique to the version in the table.
[d]The Doppler shift when the satellite is receding results in the radiation being in the protected band.

**Table A5.** Detections for Starlink V2 Mini Satellites, from tracked observations with Mopra.

| Frequency (MHz) | Total detections (# unique satellites[a]) | Weak | Strong | Interm. | Detection rate (per cent) | Total observations | ARSP 2021 footnotes[b] | Allocation |
|---|---|---|---|---|---|---|---|---|
| 1320.00[c] | 20 | 7 | 8 | 5 | 83.3 | 24 | | |
| 1350.00[c] | 22 | 3 | 14 | 5 | 95.6 | 23 | 5.149 | |
| 1440.00 | 25 | 13 | 4 | 8 | 100 | 25 | | |
| 1474.50 | 1 | 1 | – | – | 4.4 | 23 | | |
| 1620.00[c] | 5 | 1 | 3 | 1 | 83.3 | 6 | | MSS secondary (s-E) |
| 1680.00[c] | 24 | 11 | 2 | 11 | 88.9 | 27 | | |
| 2400.00 | 8 | 4 | 1 | 3 | 47.1 | 17 | | |
| 2430.00[c] | 16 | 1 | 9 | 6 | 94.1 | 17 | | |
| 2520.00 | 3 | – | – | 3 | 42.9 | 7 | | FSS primary (s-E) |
| 2578.13 | 3 | 1 | 1 | 1 | 75 | 4 | | |
| 2656.25 | 2 | 1 | – | 1 | 8.3 | 24 | 5.149, 5.208B | RAS secondary |
| 2700.00[c] | 43 (42) | 10 | 24 | 9 | 95.6 | 45 | 5.340[d] | RAS primary |
| 2760.00 | 7 | 1 | – | 6 | 87.5 | 8 | | |
| 4531.25 | 3 | 3 | – | – | 50 | 6 | | FSS primary (s-E) |
| 4650.00 | 2 | – | 1 | 1 | 40 | 5 | | FSS primary (s-E) |
| 5000.00 | 9 | 3 | 5 | 1 | 69.2 | 13 | 5.340[d] | RAS primary |
| 5072.73 | 2 | 2 | – | – | 28.6 | 7 | | |
| 5072.75 | 4 | 2 | 2 | – | 57.1 | 7 | | |
| 5156.25 | 6 | 1 | 2 | 3 | 50 | 12 | | |
| 5239.75 | 2 | 2 | – | – | 22.2 | 9 | | |
| 5241.82 | 8 | 4 | 4 | – | 88.9 | 9 | | |
| 8640.00 | 17 | 2 | 11 | 4 | 94.4 | 18 | | |

[a]Where the number of unique satellites with detections is less than the total detections.
[b]All detections in the table are also covered by AUS87 and *L*-band detections are additionally covered by AUS103.
[c]Appear in multiple tables, otherwise unique to the version in the table.
[d]The Doppler shift when the satellite is receding results in the radiation being in the protected band.

**Table A6.** Detections for Hulianwang-Digui Satellites, from tracked observations with Mopra. All detections in the table are covered by footnote AUS87, and L-band detections additionally are covered by footnote AUS103. The Doppler shift when the satellite is receding results in the radiation appearing in the 5.340 protected band.

| Frequency [MHz] | Total Detections (# unique satellites) | Weak | Strong | Interm. | Detection Rate (per cent) | Total observations | ARSP 2021 footnotes | Allocation |
|---|---|---|---|---|---|---|---|---|
| 1447.54 | 1 | – | 1 | – | 25.0 | 4 | | |
| 1448.56 | 1 | 1 | – | – | 25.0 | 4 | | |
| 1448.57 | 1 | 1 | – | – | 25.0 | 4 | | |
| 1448.90 | 1 | 1 | – | – | 25.0 | 4 | | |
| 1449.01 | 1 | 1 | – | – | 25.0 | 4 | | |
| 1449.12 | 1 | 1 | – | – | 25.0 | 4 | | |
| 1449.59 | 1 | – | 1 | – | 25.0 | 4 | | |
| 1450.44 | 1 | 1 | – | – | 25.0 | 4 | | |
| 1450.62 | 1 | – | 1 | – | 25.0 | 4 | | |
| 1450.78 | 1 | – | 1 | – | 25.0 | 4 | | |
| 1450.95 | 1 | 1 | – | – | 25.0 | 4 | | |
| 1451.06 | 1 | 1 | – | – | 25.0 | 4 | | |
| 1451.31 | 1 | 1 | – | – | 25.0 | 4 | | |
| 1451.36 | 1 | 1 | – | – | 25.0 | 4 | | |
| 1451.64 | 1 | – | 1 | – | 25.0 | 4 | | |
| 1451.96 | 1 | 1 | – | – | 25.0 | 4 | | |
| 1452.17 | 1 | 1 | – | – | 25.0 | 4 | | |
| 1452.33 | 1 | 1 | – | – | 25.0 | 4 | | |
| 1452.49 | 1 | 1 | – | – | 25.0 | 4 | | |
| 1452.65 | 1 | 1 | – | – | 25.0 | 4 | | |
| 1452.67 | 1 | 1 | – | – | 25.0 | 4 | | |
| 1452.83 | 1 | 1 | – | – | 25.0 | 4 | | |
| 1452.92 | 1 | 1 | – | – | 25.0 | 4 | | |
| 1452.99 | 1 | 1 | – | – | 25.0 | 4 | | |
| 1453.02 | 1 | 1 | – | – | 25.0 | 4 | | |
| 1453.10 | 1 | 1 | – | – | 25.0 | 4 | | |
| 1453.21 | 1 | 1 | – | – | 25.0 | 4 | | |
| 1453.28 | 1 | 1 | – | – | 25.0 | 4 | | |
| 1453.68 | 1 | – | 1 | – | 25.0 | 4 | | |
| 1455.73 | 1 | – | 1 | – | 25.0 | 4 | | |
| 1457.78 | 1 | – | 1 | – | 25.0 | 4 | | |
| 1458.24 | 1 | 1 | – | – | 25.0 | 4 | | |
| 1458.44 | 1 | 1 | – | – | 25.0 | 4 | | |
| 1458.47 | 1 | 1 | – | – | 25.0 | 4 | | |
| 1458.61 | 1 | – | – | – | 25.0 | 4 | | |
| 1458.63 | 1 | 1 | – | – | 25.0 | 4 | | |
| 1458.79 | 1 | – | 1 | – | 25.0 | 4 | | |
| 1458.88 | 1 | – | – | 1 | 25.0 | 4 | | |
| 1458.92 | 1 | 1 | – | – | 25.0 | 4 | | |
| 1458.97 | 1 | – | – | 1 | 25.0 | 4 | | |
| 1459.06 | 1 | – | – | 1 | 25.0 | 4 | | |
| 1459.35 | 1 | 1 | – | – | 25.0 | 4 | | |
| 1459.82 | 1 | – | 1 | – | 25.0 | 4 | | |
| 1460.29 | 1 | 1 | – | – | 25.0 | 4 | | |
| 1460.51 | 1 | 1 | – | – | 25.0 | 4 | | |
| 1460.66 | 1 | 1 | – | – | 25.0 | 4 | | |
| 1460.67 | 1 | 1 | – | – | 25.0 | 4 | | |
| 1503.07 | 2 | 1 | – | 1 | 50.0 | 4 | | |
| 1504.70 | 2 | 1 | – | 1 | 50.0 | 4 | | |
| 1505.09 | 2 | 1 | – | 1 | 50.0 | 4 | | |
| 1506.89 | 2 | 1 | – | 1 | 50.0 | 4 | | |
| 1767.74 | 1 | – | 1 | – | 12.5 | 8 | | |
| 1769.79 | 1 | – | 1 | – | 12.5 | 8 | | |
| 1771.84 | 1 | – | 1 | – | 12.5 | 8 | | |
| 1773.88 | 1 | – | 1 | – | 12.5 | 8 | | |
| 1775.93 | 1 | – | 1 | – | 12.5 | 8 | | |
| 1777.97 | 1 | – | 1 | – | 12.5 | 8 | | |
| 1780.02 | 1 | – | 1 | – | 12.5 | 8 | | |
| 1782.07 | 1 | – | 1 | – | 12.5 | 8 | | |
| 1784.11 | 1 | – | 1 | – | 12.5 | 8 | | |

**Table A6** – *continued*

| Frequency [MHz] | Total Detections (# unique satellites) | Weak | Strong | Interm. | Detection Rate (per cent) | Total observations | ARSP 2021 footnotes | Allocation |
|---|---|---|---|---|---|---|---|---|
| 1786.16 | 1 | – | 1 | – | 25.0 | 4 | | |
| 2349.83 | 1 | – | – | 1 | 16.7 | 6 | | |
| 2379.50 | 1 | – | – | 1 | 16.7 | 6 | | |
| 2380.00 | 1 | – | – | 1 | 16.7 | 6 | | |
| 2380.04 | 1 | – | – | 1 | 16.7 | 6 | | |
| 2381.54 | 1 | 1 | – | – | 16.7 | 6 | | |
| 2383.59 | 2 | 2 | – | – | 33.3 | 6 | | |
| 2385.64 | 2 | 2 | – | – | 33.3 | 6 | | |
| 2386.98 | 2 | – | – | 2 | 33.3 | 6 | | |
| 2387.68 | 2 | – | 2 | – | 33.3 | 6 | | |
| 2388.98 | 2 | – | – | 2 | 33.3 | 6 | | |
| 2389.73 | 2 | – | 2 | – | 33.3 | 6 | | |
| 2389.98 | 2 | – | – | 2 | 33.3 | 6 | | |
| 2390.04 | 1 | – | – | 1 | 16.7 | 6 | | |
| 2390.98 | 2 | – | – | 2 | 33.3 | 6 | | |
| 2391.77 | 2 | – | 2 | – | 33.3 | 6 | | |
| 2392.98 | 2 | – | – | 2 | 33.3 | 6 | | |
| 2393.82 | 2 | – | 2 | – | 33.3 | 6 | | |
| 2393.98 | 1 | – | – | 1 | 16.7 | 6 | | |
| 2395.38 | 1 | 1 | – | – | 16.7 | 6 | | |
| 2395.86 | 1 | – | 1 | – | 16.7 | 6 | | |
| 2395.87 | 2 | – | 2 | – | 33.3 | 6 | | |
| 2396.89 | 1 | 1 | – | – | 16.7 | 6 | | |
| 2396.91 | 1 | 1 | – | – | 16.7 | 6 | | |
| 2397.43 | 1 | 1 | – | – | 16.7 | 6 | | |
| 2397.44 | 1 | 1 | – | – | 16.7 | 6 | | |
| 2397.52 | 1 | – | 1 | – | 16.7 | 6 | | |
| 2397.91 | 2 | – | 2 | – | 33.3 | 6 | | |
| 2399.96 | 3 | – | 3 | – | 50.0 | 6 | | |
| 2401.05 | 3 | – | – | 3 | 50.0 | 6 | | |
| 2401.69 | 1 | – | – | 1 | 16.7 | 6 | | |
| 2401.71 | 1 | – | 1 | – | 16.7 | 6 | | |
| 2401.73 | 2 | 2 | – | – | 33.3 | 6 | | |
| 2401.74 | 1 | – | 1 | – | 16.7 | 6 | | |
| 2402.00 | 3 | – | 3 | – | 50.0 | 6 | | |
| 2404.05 | 3 | – | 3 | – | 50.0 | 6 | | |
| 2405.05 | 3 | – | 2 | 1 | 50.0 | 6 | | |
| 2406.09 | 2 | – | 2 | – | 33.3 | 6 | | |
| 2407.05 | 2 | – | 1 | 1 | 33.3 | 6 | | |
| 2408.14 | 2 | – | 2 | – | 33.3 | 6 | | |
| 2410.19 | 3 | – | 3 | – | 50.0 | 6 | | |
| 2412.23 | 2 | – | 2 | – | 33.3 | 6 | | |
| 2414.28 | 2 | – | 2 | – | 33.3 | 6 | | |
| 2415.12 | 2 | – | – | 2 | 33.3 | 6 | | |
| 2416.33 | 2 | – | 2 | – | 33.3 | 6 | | |
| 2417.12 | 2 | – | – | 2 | 33.3 | 6 | | |
| 2418.12 | 2 | – | – | 2 | 16.7 | 12 | | |
| 2418.37 | 2 | – | 2 | – | 16.7 | 12 | | |
| 2536.51 | 1 | – | 1 | – | 8.3 | 12 | | |
| 2536.66 | 1 | – | 1 | – | 8.3 | 12 | | |
| 2536.81 | 1 | 1 | – | – | 8.3 | 12 | | |
| 2536.92 | 1 | 1 | – | – | 8.3 | 12 | | |
| 2536.94 | 1 | 1 | – | – | 8.3 | 12 | | |
| 2537.04 | 1 | – | 1 | – | 8.3 | 12 | | |
| 2537.14 | 1 | 1 | – | – | 8.3 | 12 | | |
| 2537.16 | 1 | 1 | – | – | 8.3 | 12 | | |
| 2537.27 | 1 | 1 | – | – | 8.3 | 12 | | |
| 2537.42 | 1 | – | 1 | – | 8.3 | 12 | | |
| 2537.57 | 1 | – | 1 | – | 8.3 | 12 | | |
| 2537.79 | 1 | 1 | – | – | 8.3 | 12 | | |
| 2670.03 | 1 | 1 | – | – | 20.0 | 5 | | RAS secondary |

**Table A6** – *continued*

| Frequency [MHz] | Total Detections (# unique satellites) | Weak | Strong | Interm. | Detection Rate (per cent) | Total observations | ARSP 2021 footnotes | Allocation |
|---|---|---|---|---|---|---|---|---|
| 2703.79 | 2 | 2 | – | – | 40.0 | 5 | | |
| 2707.88 | 2 | 1 | 1 | – | 40.0 | 5 | | |
| 2709.93 | 4 | – | 4 | – | 40.0 | 10 | | |
| 2711.97 | 4 | – | 4 | – | 40.0 | 10 | | |
| 2714.02 | 2 | – | 2 | – | 40.0 | 5 | | |
| 2716.06 | 1 | – | 1 | – | 20.0 | 5 | | |
| 2716.07 | 2 | – | 2 | – | 40.0 | 5 | | |
| 2718.11 | 2 | – | 2 | – | 40.0 | 5 | | |
| 2720.16 | 2 | – | 2 | – | 40.0 | 5 | | |
| 2722.20 | 2 | – | 2 | – | 40.0 | 5 | | |
| 2724.25 | 2 | – | 2 | – | 40.0 | 5 | | |
| 2726.30 | 2 | – | 2 | – | 40.0 | 5 | | |
| 2728.34 | 2 | – | 2 | – | 40.0 | 5 | | |
| 2730.39 | 2 | – | 2 | – | 40.0 | 5 | | |
| 2732.43 | 2 | – | 2 | – | 40.0 | 5 | | |
| 2734.48 | 2 | – | 2 | – | 40.0 | 5 | | |
| 2736.52 | 2 | – | 2 | – | 40.0 | 5 | | |
| 2738.57 | 2 | – | 2 | – | 40.0 | 5 | | |
| 2740.62 | 2 | – | 2 | – | 40.0 | 5 | | |
| 4700.00 | 1 | – | 1 | – | 50 | 2 | | FSS primary (s-E) |
| 4999.96 | 1 | – | 1 | – | 100 | 1 | 5.340 | RAS primary |

[a]Where the number of unique satellites with detections is less than the total detections.
[b]All detections in the table are also covered by AUS87 and *L*-band detections are additionally covered by AUS103.
[c]Appear in multiple tables, otherwise unique to the version in the table.
[d]The Doppler shift when the satellite is receding results in the radiation being in the protected band.

This paper has been typeset from a TEX/LATEX file prepared by the author.